\documentclass[%
 reprint,
 amsmath,amssymb,
 aps,
]{revtex4-2}

\usepackage{hyperref}
\hypersetup{
    colorlinks=true, 
    linkcolor=cyan,
    citecolor=magenta, 
    filecolor=magenta, 
    urlcolor=cyan,
    runcolor=cyan
}
\usepackage{hyperref}
\usepackage{graphicx}
\usepackage{dcolumn}
\usepackage{bm}
\usepackage{subfig}
\usepackage{color,soul}
\usepackage{float}
\usepackage{mathtools}
\usepackage{siunitx}
\usepackage{booktabs}

\usepackage{tikz}
\usetikzlibrary{shapes,arrows}

\usepackage[english]{babel}

\makeatletter
\renewcommand{\fnum@figure}{FIG. \thefigure}
\makeatother

\begin{document}

\title{Reconstructing Gamma-ray Energy Distributions from PEDRO Pair Spectrometer Data}

\author{M. Yadav$^{1,2,3, \alpha}$}
\email{yadavmonika@g.ucla.edu}
\author{M. H. Oruganti$^{1, \alpha}$}
\email{maanashemanth22@g.ucla.edu}
\thanks{$^{\alpha}$These two authors contributed equally}
\author{B. Naranjo$^{1}$}
\author{G. Andonian$^{1}$}
\author{{\"O.} Apsimon$^{2,3}$}
\author{C. P. Welsch$^{2,3}$}
\author{J. B. Rosenzweig$^{1}$}

\affiliation{
$^1$Department of Physics and Astronomy, University of California Los Angeles, California 90095, USA
}
 \affiliation{ 
 $^2$ Cockcroft Institute, Warrington WA4 4AD, UK}
\affiliation{ 
 $^3$Department of Physics, University of Liverpool, Liverpool L69 3BX, UK}

\date{\today}
\begin{abstract} 

Photons emitted from high-energy electron beam interactions with high-field systems, such as the upcoming FACET-II experiments at SLAC National Accelerator Laboratory, may provide deep insight into the electron beam's underlying dynamics at the interaction point. With high-energy photons being utilized to generate electron-positron pairs in a novel spectrometer, there remains a key problem of interpreting the spectrometer's raw data to determine the energy distribution of the incoming photons. This paper uses data from simulations of the primary radiation emitted from electron interactions with a high-field, short-pulse laser to determine optimally reliable methods of reconstructing the measured photon energy distributions. For these measurements, recovering the emitted 10 MeV to 10 GeV photon energy spectra from the pair spectrometer currently being commissioned requires testing multiple methods to finalize a pipeline from the spectrometer data to incident photon and, by extension, electron beam information. In this study, we compare the performance QR decomposition, a matrix deconstruction technique and neural network with and without maximum likelihood estimation (MLE). Although QR decomposition proved to be the most effective theoretically, combining machine learning and MLE proved to be superior in the presence of noise, indicating its promise for analysis pipelines involving high-energy photons. 
\end{abstract}

\maketitle

\section{Introduction}

The pair spectrometer designed for strong-field quantum electrodynamics regime (SFQED) experiments at FACET-II addresses the need for precise measurements of high-energy photon spectra ranging from 10 MeV to 10 GeV \cite{Matheron2023}. Current methods, such as $\gamma$-ray calorimetry and differential $\gamma$-ray absorption, face significant challenges due to pile-up and ambiguous spectral information in high-flux environments. The pair spectrometer overcomes these limitations by employing a modular magnetic geometry that analyzes positron-electron pairs produced in a conversion target. Its design features a symmetric configuration to record electron and positron responses, making it suitable for high-energy (up to 10 GeV) photon measurements. The integrated solution builds on established magnetic analysis techniques, like Enge's achromatic mirror \cite{enge1963achromatic}, and is tailored to meet the extreme demands of SFQED environments, where traditional diagnostics fall short. While the pair spectrometer provides a major experimental hardware upgrade, the software to interpret these measurements effectively remains unknown until now. 

Broadly speaking, experiments studying the behavior of electron beams in advanced accelerator contexts can be divided into two classes: characterization of beam properties themselves and investigations of their interactions with external factors. The latter category can be further divided, in the context of the broad FACET-II program, into high-energy electron interactions with dense plasmas and those with high-intensity laser beams. For example, the FACET-II experiments on beam filamentation \cite{filamentation_1,filamentation_2} phenomena in dense plasma utilize very large fields that emit high energy photons \cite{Zampetakis}. The highest energy photons produced in FACET-II experiments are obtained from nonlinear inverse Compton scattering (NLCS) where the electron beam encounters ultra-fast, very intense laser beams. In the SFQED regimen, interactions result in multiple photons being absorbed and released as a single, higher energy photon \cite{di2018implementing}. The aims of this experiment require very detailed spectral measurement from tens of MeV through the full energy of the electron beam.  

The upcoming installation of the UCLA Positron-Electron Detector for Radiative Observations (PEDRO) at the FACET-II facility marks a significant enhancement of experimental capabilities.
It is essential for studying high-energy, high-field interactions, where the photon energy exceeds that appropriate for use in another UCLA-built spectrometer based on Compton scattering due to the onset of pair creation. PEDRO has an innovative design capable of spanning the needed spectral energy reach. 

\begin{figure*}[t]  {\includegraphics[width=\textwidth]{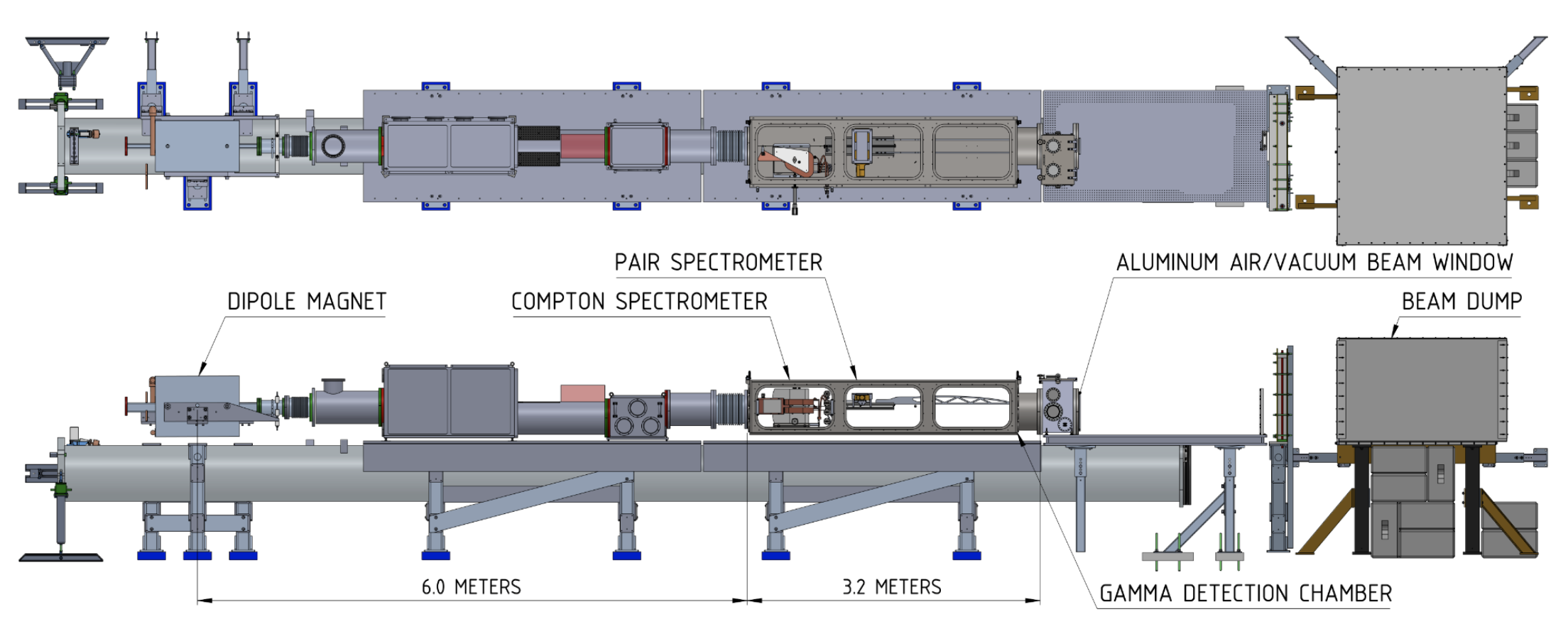}}
  \caption{A schematic map of the FACET-II beamline at dump \cite{PEDRO-hardware}. The beam travels from left to right, the interaction point is nearly 13 meters from the spectrometer. Six meters downstream of the dipole, the gamma detection chamber (GDC) is positioned along the beamline. The first third of the GDC contains the Compton spectrometer, while the remaining two-thirds house the pair spectrometer. Beyond the GDC, the beam proceeds to the dump table, which is equipped with various diagnostics.}
  \label{fig:GDC_1}
\end{figure*}

Beyond this application, PEDRO measurements represent a new tool for the measurement of broad range, directed $\gamma$-ray beams. These beams are crucial for exploring a wide array of physical phenomena and have promising practical applications. High-flux, high-energy laboratory $\gamma$-ray sources particularly in the case of inverse Compton scattering in an intense laser field. These sources are essential for studying not only SFQED phenomena -- through the emission process itself -- but also nuclear phenomena in a burgeoning field termed nuclear photonics. 

In the SFQED case, quantum effects, such as quantum radiation reaction, stochastic photon emission, and electron-positron pair production, significantly influence the spectrum and angular distribution of photons emitted during Compton scattering of ultra-relativistic electrons in intense laser fields. 
The development of detectors capable of precisely measuring such spectra is crucial for advancing our understanding of SFQED.
This need justifies the inclusion of $\gamma$-ray spectrometers in large-scale SFQED experiments, such as E-320 at SLAC. 

The PEDRO $\gamma$-ray spectrometer faces particular experimental challenges in resolving photon beams that are both high-flux and high-energy. These challenges lead to the demand for a robust analysis method that can reproduce the primary $\gamma$-ray spectrum incident on the spectrometer. For FACET-II, PEDRO utilizes a thin beryllium wire that interacts with such photons via its strong nuclear field to produce positron-electron pairs. Each pair has the same total energy as the incident higher energy photons; however, the individual energies of the positron and the electrons are not evenly distributed. Under the influence of a magnetic field, the initially highly forward-directed electrons and positrons bend in opposite directions and at differing energy-dependent angles. Those with higher energies travel proportionally further down the spectrometer before striking a detector cell in the GDC. This array of detection cells measures the number of electrons or positrons that strike at the full range of relevant locations. This information can directly establish the energy of one incoming photon. However, the analysis becomes nontrivial when dealing with billions of photons over a large energy range.
 
The spectrometer's detection of any given $\gamma$-ray distribution can be modeled through a single response matrix (this matrix was derived from a Monte Carlo simulation of tens of thousands of photon distributions). This then reduces the problem to inverting a matrix, while additionally permitting accounting for the presence of noise in the detection chamber. However, given that this matrix is not square, and therefore not trivially invertible, this introduces complications to the problem that demand some more deliberate methods of approach.

The initial consideration should be to establish a gold standard, which can then serve as a benchmark for comparing alternative reconstruction methods. Experimental noise adds several unpredictable layers of complexity to the process of reconstruction, but if the system is treated as ideal for a moment, an arsenal of linear algebraic and matrix analysis techniques becomes relevant to establishing a baseline. QR decomposition, a method of splitting a rectangular matrix into orthonormal and diagonal components \cite{Gander2003AlgorithmsFT}, stands out as a clear-cut analysis method for generating a linear system that can consistently be resolved to yield a unique solution. These properties are further explored in Sec. \ref{QR}.

\begin{figure}[hbt!]
  \centering{\includegraphics[width=0.47\textwidth]{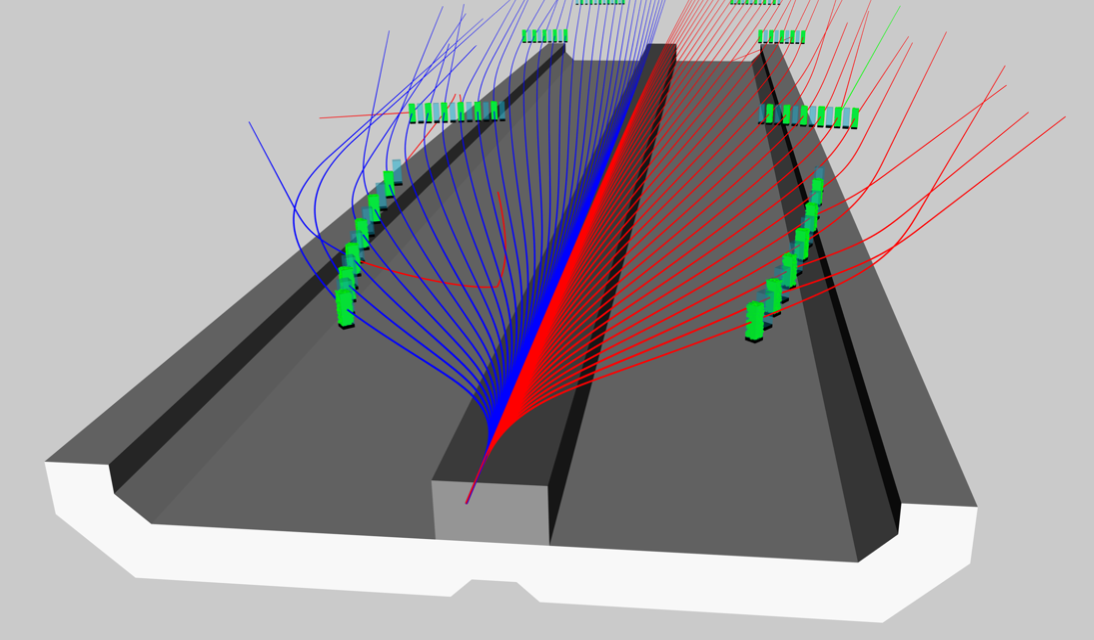}}
  \caption{Cutaway view of design trajectories. Electrons are shown in red; positrons in blue, and photons in green with energies span 10 MeV through 10 GeV. The Cherenkov cells are oriented normally to incoming design trajectories.}
  \label{fig:pedro}
\end{figure}

With a gold standard selected, the methods that can resist noise must be explored to ensure the reconstruction is robust. Machine learning (ML) techniques have been employed in several innovative ways within the context of advanced accelerator research, particularly at the FACET-II facility. These applications have a variety of aims, including enhancing beam quality, simplifying experimental setups, and predicting key parameters for intricate wakefield acceleration experiments. These applications include the development of virtual diagnostics \cite{Claudio} or digital twins. ML has also been used to reconstruct beam phase space distributions from experimental data \cite{A.wolski,Scheinker2021}. Recent studies have highlighted the power and versatility of ML methods used in particle accelerators as tools for prediction, control, and optimization of accelerator performance \cite{Sanchez-Gonzalez2017-fl-nature-2017,edelen2016first,2016_Edelen, A_Scheinker_2018}. 

Given these previous applications of machine learning in accelerator physics, it stood to reason to consider applying a neural network to this problem as well. Since experimentation often introduces noise to the data, a neural network could be trained to withstand the effects of noise while recovering the relationship between PEDRO readouts and incoming photon energy distributions. By extension, this problem also proved to be iterable through the Sheep-Vardi algorithm, allowing for the testing of maximum-likelihood estimation (MLE) as another possible method of reconstruction.

The last major aspect to consider is the presence of noise. PEDRO itself presents with Poisson noise in the readout when detecting any spectrum, no matter how clean, due to interactions within the Cherenkov cells that detect the electrons and positrons. Additionally, background X-rays are inevitable in a facility such as FACET-II, and this noise can be modeled with a normal distribution that has a standard deviation of 5\% of its mean, discussed elsewhere in a spectrometer design paper by Brian Naranjo et al. 

The paper is organized as follows. In Sec. \ref{sec:Pairspec}, we discuss possible available analysis methods of QR decomposition, ML, and an ML-MLE combination. In Sec. \ref{sec:discussion}, the results are compared across the three methods. Finally, we offer conclusions and discuss the outlook for future work in Sec. \ref{sec:conclusion}.

\section{Methods}
\label{sec:Pairspec}

For the PEDRO pair spectrometer, the need for a robust method for determining the initial photon energy distribution becomes apparent when one considers a monochromatic $\gamma$-ray source; the resultant pair distributions are broad-range. As such, when considering electron-positron pairs produced from primary photons, it is key to recognize the linear relationship between the energy distribution of photons and the spectrometer's response. This linear relationship can be modeled in Eq: \ref{eq:19} as follows:

\begin{equation}
    \begin{cases}
      R_{1,1}*x_1 + ... + R_{1,64}*x_{64} = y_1 \\
      R_{2,1}*x_1 + ... + R_{2,64}*x_{64} = y_2 \\
      ... \\
      R_{128,1}*x_1 +  ... + R_{128,64}*x_{64} = y_{128} \\
    \end{cases}
     \label{eq:19}
\end{equation}

where the vector ($x_1$, $x_2$,..., $x_{64}$) represents the gamma energy distribution, $R_{i,j}$ are scalar coefficients, and the elements ($y_1$, $y_2$,..., $y_{128}$) represent the spectrometer's responses. More specifically, $x_i$ represents the number of photons in each logarithmically-spaced energy bin; for example, $x_1$ = 10 indicates that there are 10 photons with energies between 10 MeV and 11.18 MeV in the incoming $\gamma$-rays. Additionally, $y_j$ represents the number of electrons or positrons that interacted in the spectrometer's $j^{th}$ detector. 

In the above system of equations, and throughout the rest of this paper, the \textit{x}-vector will refer to the original distribution of photons based on their energies. While the energy values will become more relevant when interpreting the model's output, the numerical values that split the energy distribution into logarithmic bins were chosen for convenience and did not impact the model's logic or construction. The \textit{y}-vector will always refer to the electron-positron spectrum that PEDRO outputs in response to the corresponding \textit{x}-vector, or incoming photon energy distribution. 

Additionally, this linear relationship permits building a matrix that can be used to generate any number of training and test cases. These cases are instrumental for training and evaluating the precision of various models in the task of recovering photon energy distributions from spectral responses. The following matrix with coefficients $R_{i,j}$ will be referred to as the response matrix $R$ since it models how the spectrometer will respond to a given photon energy distribution.

\begin{equation} \label{eq:pedro}
    \begin{bmatrix} 
        R_{1,1} & R_{1,1} & \cdots & R_{1,64} \\
        R_{2,1} & R_{2,1} & \cdots & R_{2,64} \\
        \vdots       & \vdots        & \ddots &  \vdots \\
        R_{128,1} & R_{128,1} & \cdots & R_{128,64} \\
    \end{bmatrix}
    \times
    \left[
        \begin{array} {c} 
            x_1 \\ 
            x_2 \\ 
            \vdots \\ 
            x_{64} 
        \end{array}
    \right] 
    = 
    \left[ 
        \begin{array}{c}
            y_1 \\ 
            y_2 \\ 
            \vdots \\ 
            y_{128} 
        \end{array} 
    \right]
\end{equation}

To account for inherent readout noise, when the \textit{y}-vector was calculated, for every element in the vector, it was recalculated based on picking from a Poisson distribution centered at that original detected value. For example, if there were 1e6 positrons generated in the first slot of the vector, then a random point would be selected from a Poisson distribution (say 1.8e05) with a mean of 1e6 to replace that initial value of 1e6. 

Additionally, to account for background x-rays, the new values calculated from the Poisson distribution were then established as means for their respective normal distributions, each with a standard deviation of 5\% of those new values. Continuing from the previous example, taking 1.8e05 as the initial value, a normal distribution with $\mu$ = 1.8e05 and $\sigma$ = 0.05 * 1.8e05 = 9e03 would be generated to sample a single point. If 173213.3639 was generated, then the integer count (173213) would be recorded as the number of hits detected in that slot. The vector with the values from layered noise would then be fed as the inputs for the various reconstruction methods. 

The application of QR decomposition to the \textit{R} matrix, the development of a machine-learning model, and the combination of the ML model with an MLE algorithm were analyzed to determine the best method of recovering the \textit{x}-vector given a \textit{y}-vector. To determine the efficacy of each approach, there were five standard test cases to which each method was applied. They are listed as follows:

\begin{itemize}
  \item Mono-energetic: The original spectrum contained $10^8$ photons exclusively in one unique bin and 0 photons elsewhere. Noise with a level of 100 was added from a Poisson distribution to introduce experimentally relevant variability to the spectrum. 
  \item Arbitrary case: The original spectrum contained only a random number of photons between 0 and $10^{10}$ in all of the photons. 
  \item Three other smooth cases were derived from experimental scenarios (nonlinear Compton scattering, strong field quantum electrodynamics, and filamentation in plasma).
\end{itemize}

\noindent The frequency bin for the mono-energetic case was chosen at random. For each of these test cases, and in future experiments, the spectrometer's ability to reconstruct frequencies that were resolved within a factor $10^{-4}$ from the spectrum's peak was considered. It is noted that in any given spectra, this narrow range alone can be reliably reconstructed even with the interference of measurement noise. This lower noise limit is indicated in each of the following figures for non-monochromatic spectra by a red dashed line, and any behavior of measured spectra below that line was not used to determine the reconstruction method's efficacy.

\subsection{QR Decomposition}\label{QR}

The computational analysis goal is to invert the response matrix \textit{R} effectively so, given a PEDRO spectrum, the originating \textit{x}-vector may be recovered. This recovery method required no additional information beyond the matrix to be decomposed (in this case, \textit{R}). 

\begin{equation}
\label{eq:QReqn}
    \hat{R} = \hat{Q} * \hat{S}
\end{equation}

In this decomposition, $\hat{Q}$ is a square orthonormal matrix, meaning each of the column vectors $Q_i$ are orthogonal to each other and have a generalized unity length. This yields a convenient property that the transpose of $\hat{Q}$, $\hat{Q}^T$, is its inverse matrix, as shown below:

\begin{equation}
    \hat{Q} * \hat{Q}^T 
    =
   \begin{bmatrix} 
        \vdots & \vdots & \cdots & \vdots \\
         Q_1 & Q_2 & \cdots & Q_{64} \\
        \vdots & \vdots & \cdots & \vdots \\
    \end{bmatrix}
    \times
    \begin{bmatrix} 
        \cdots & Q_1 & \cdots \\
        \cdots & Q_2 & \cdots \\
        \vdots & \vdots & \vdots \\
        \cdots & Q_{64} & \cdots \\
    \end{bmatrix}
\end{equation}

\begin{equation}
     =
    \begin{bmatrix} 
        <Q_1,Q_1> & \cdots & <Q_1,Q_{64}> \\
        <Q_2,Q_1> &  \cdots & \vdots \\
        \vdots & \ddots & \vdots \\
        <Q_{64},Q_1> & \cdots & <Q_{64},Q{64}> \\ 
    \end{bmatrix}
\end{equation}

\begin{equation}
    =
    \begin{bmatrix} 
        1 & 0 & \cdots & 0 \\
        \vdots & \vdots & \ddots & \vdots \\
        0 & 0 & \cdots & 1 \\ 
    \end{bmatrix}
\end{equation}

As such, using this property and substituting the decomposition into the matrix equation that models the spectrometer's response, the equation transforms into:

\begin{equation}
    y = \hat{R}x = (\hat{Q}\hat{S})*x
    => \hat{Q}^Ty = \hat{Q}^T\hat{Q} * \hat{S}x
\end{equation}

\begin{equation} \label{eq:optimizer}
    => \hat{S}x = \hat{Q}^Ty = b
\end{equation}

Since, for a given response or \textit{y}-vector, the term $\hat{Q}^Ty$ will be constant, it will be referred to as the \textit{b}-vector throughout the remainder of this discussion. Under these assumptions, the method for finding the \textit{x}-vector is reduced to determining which solution will minimize the following function:

\begin{equation} 
    f(x) = ||\hat{S}x - b||^2
\end{equation}

Minimizing $f(x)$ will result in a solution for Eq.~\ref{eq:optimizer}, meaning using the least-squares optimization algorithm will provide the most likely energy distribution. Before implementing this algorithm, the matrix $\hat{S}$ values must be accounted for to determine whether the function is strongly convex. Strong convexity means the function is guaranteed to have a unique global minimum, which ensures the least-squares algorithm will converge to only one possible answer for a given \textit{b}-vector. 

To confirm strong convexity, the second derivative of $f(x)$ must generate a positive definite matrix, This condition is equivalent to stating that the representation must be a symmetric matrix with strictly positive eigenvalues (all of the eigenvalues $\lambda_i > \;$0).

\begin{equation}
    \frac{df(x)}{dx} = 2\hat{S}^T(\hat{S}x - b) = 2\hat{S}^T\hat{S}x - 2\hat{S}^Tb
\end{equation}

\begin{equation}
    => \frac{d^2f(x)}{dx^2} = 2\hat{S}^T\hat{S}
\end{equation}

Since $\frac{d^2f(x)}{dx^2}$ is equal to the product of a matrix with its transpose, it follows that:

\begin{equation}
\label{eq:Symeqn}
    (2\hat{S}^T\hat{S})^T = 2(\hat{S}^T)(\hat{S}^T)^T = 2\hat{S}^T\hat{S}
\end{equation}

\begin{figure}[t]
   \centering
     \subfloat[Mono-energetic case]{\includegraphics[width=\columnwidth]{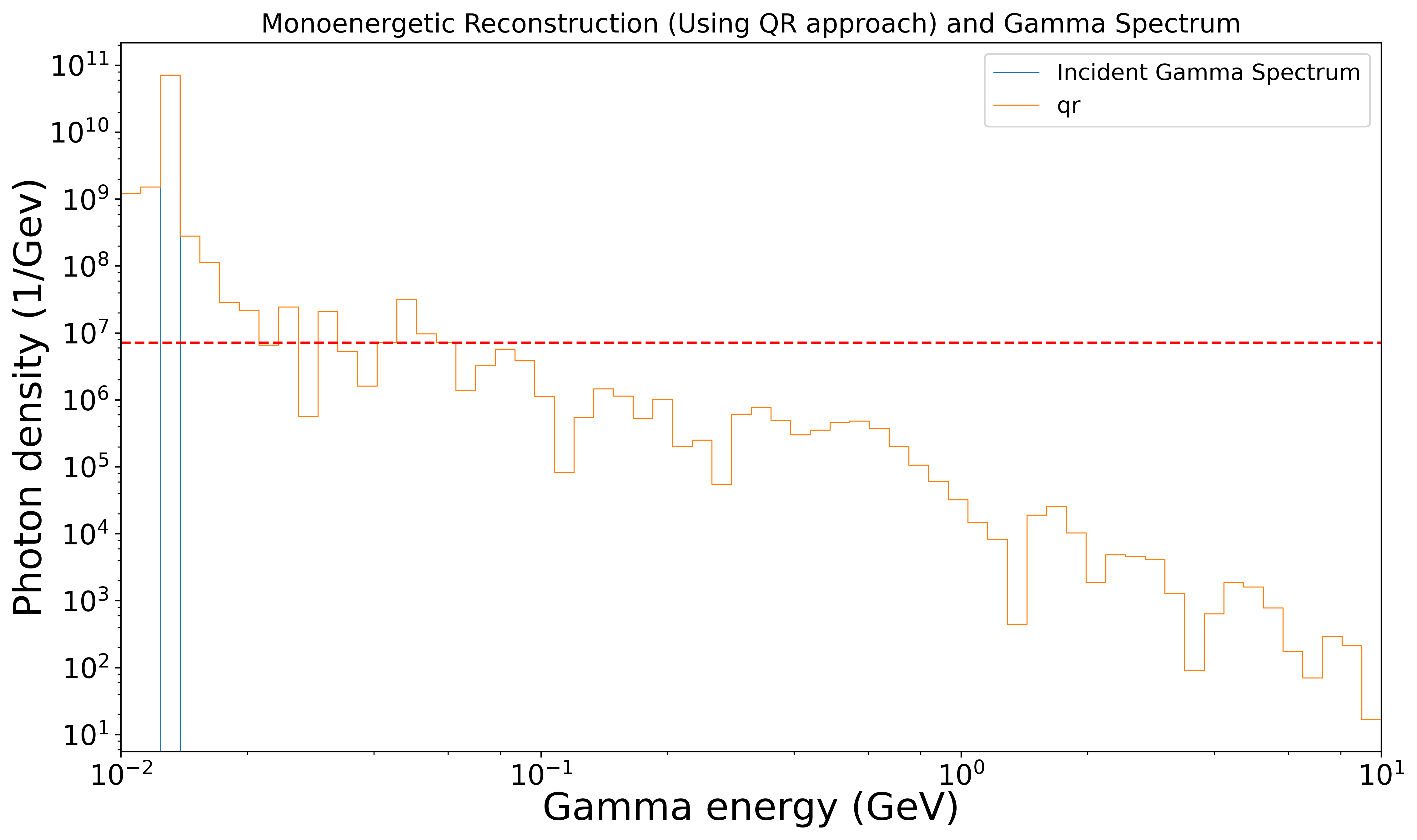}}\\
      \subfloat[Arbitrary discrete case]{\includegraphics[width=\columnwidth]{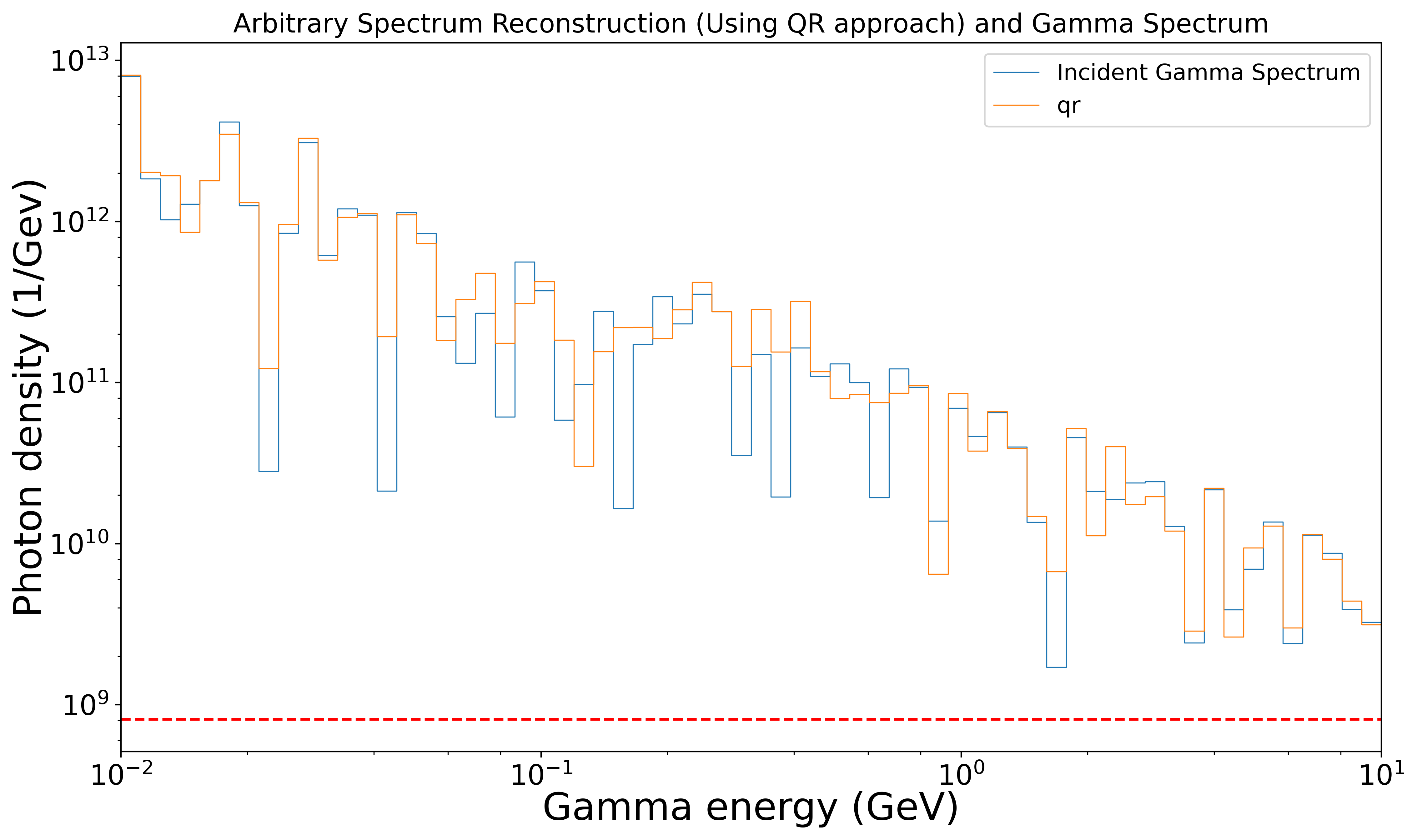}}
   \caption{Discrete reconstructed $\gamma$-ray energy distributions using QR decomposition.}
  \label{fig:6}
\end{figure}

Using Eq.~\ref{eq:Symeqn}, $\frac{d^2f(x)}{dx^2}$ forms a symmetric matrix, meaning the only test that remains to determine the definite positive nature of $\frac{d^2f(x)}{dx^2}$ is to determine its eigenvalues. The Python NumPy library contains a function that takes a matrix as input and returns an array of its eigenvalues. Applying this function to $\frac{d^2f(x)}{dx^2}$ demonstrated that it indeed does have only strictly positive eigenvalues.

Thus, using the least squares optimization algorithm will avoid the issue of converging on potentially multiple solutions. After generating the QR decomposition, like the other approaches, this algorithm was tested against both discrete cases and more continuous cases.

\subsubsection{Establishing the Baseline}\label{QR_compare}

\begin{figure}[hbt!]
  \centering
    \subfloat[Ideal Reconstruction]{\includegraphics[width=\columnwidth]{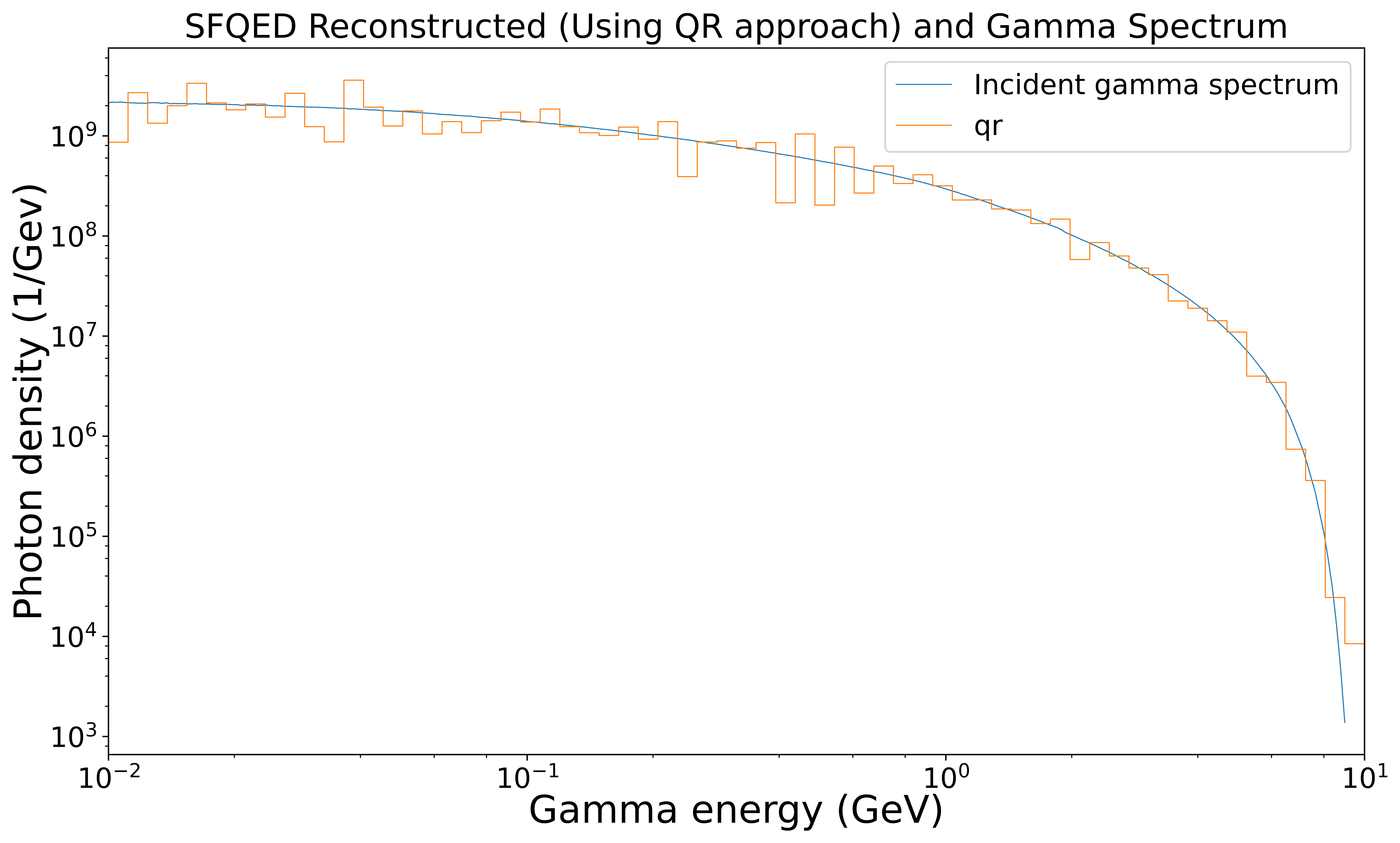}}\\
    \subfloat[Noisy Reconstruction]{\includegraphics[width=\columnwidth]{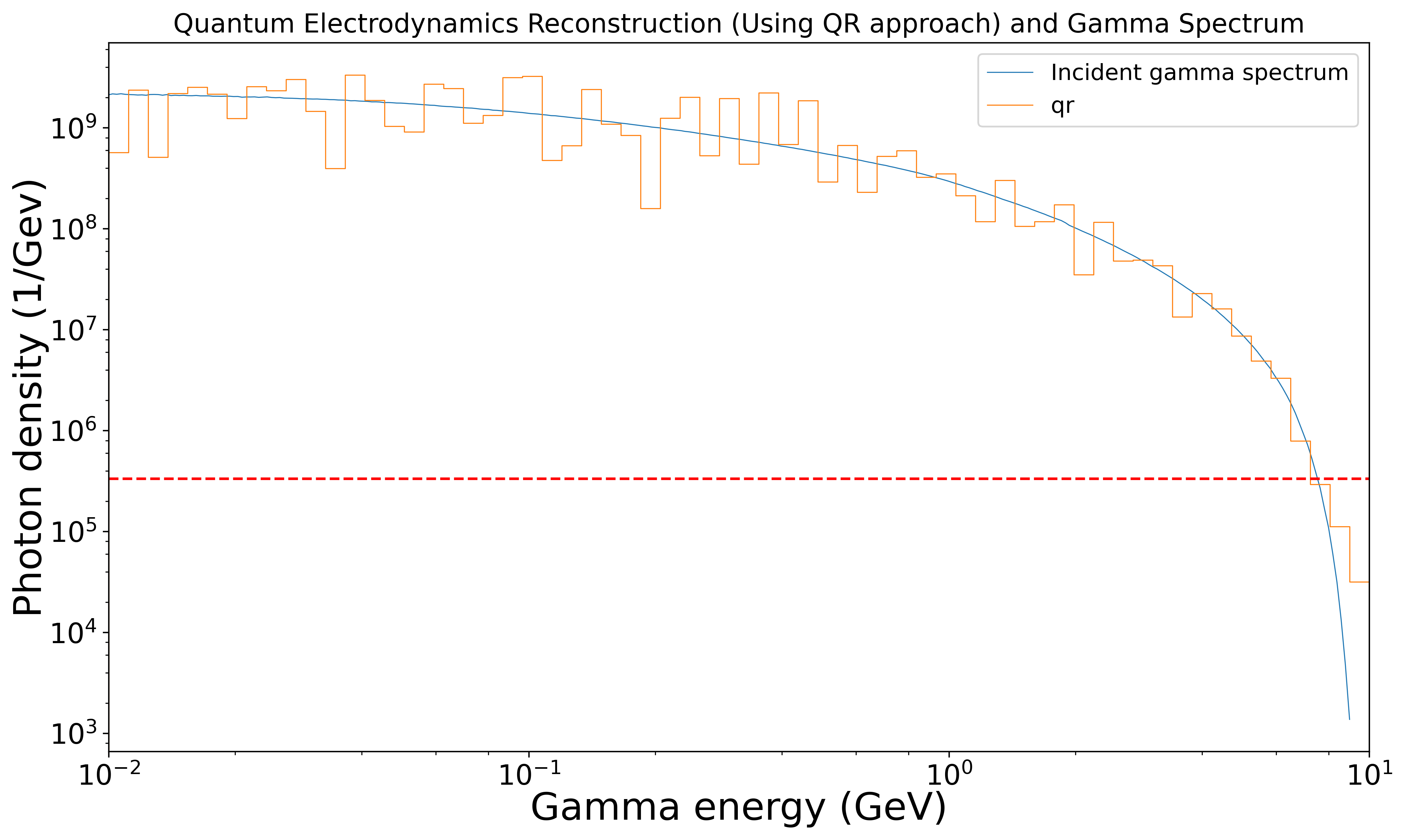}}\\
  \caption{Smooth reconstructed $\gamma$-ray distributions using QR decomposition.}
  \label{fig:QR_comparison}
\end{figure}

QR decomposition is a method that relies on decomposing a strictly linear system, and the presence of nonlinear noise will likely inhibit this method's reconstructive capabilities. Before applying noise to all cases, this method was evaluated on one experiment, SFQED, with and without noise to investigate how spectrometer readouts could impact the reconstruction process. Fig. \ref{fig:QR_comparison} compares the reconstruction with and without noise, and the case without noise results in a reconstruction that follows the true spectrum more closely. Though this was expected, it helps establish that the gold standard for matrix reconstruction clearly diminishes in its reconstructive power as real-world noise becomes more prominent. Therefore, there is a need to examine this method in other contexts and compare it to more robust methods that can adapt to this noise issue. 

\subsubsection{Assessing the method}

\begin{figure}[hbt!]
  \centering
    \subfloat[Nonlinear Compton scattering case]{\includegraphics[width=\columnwidth]{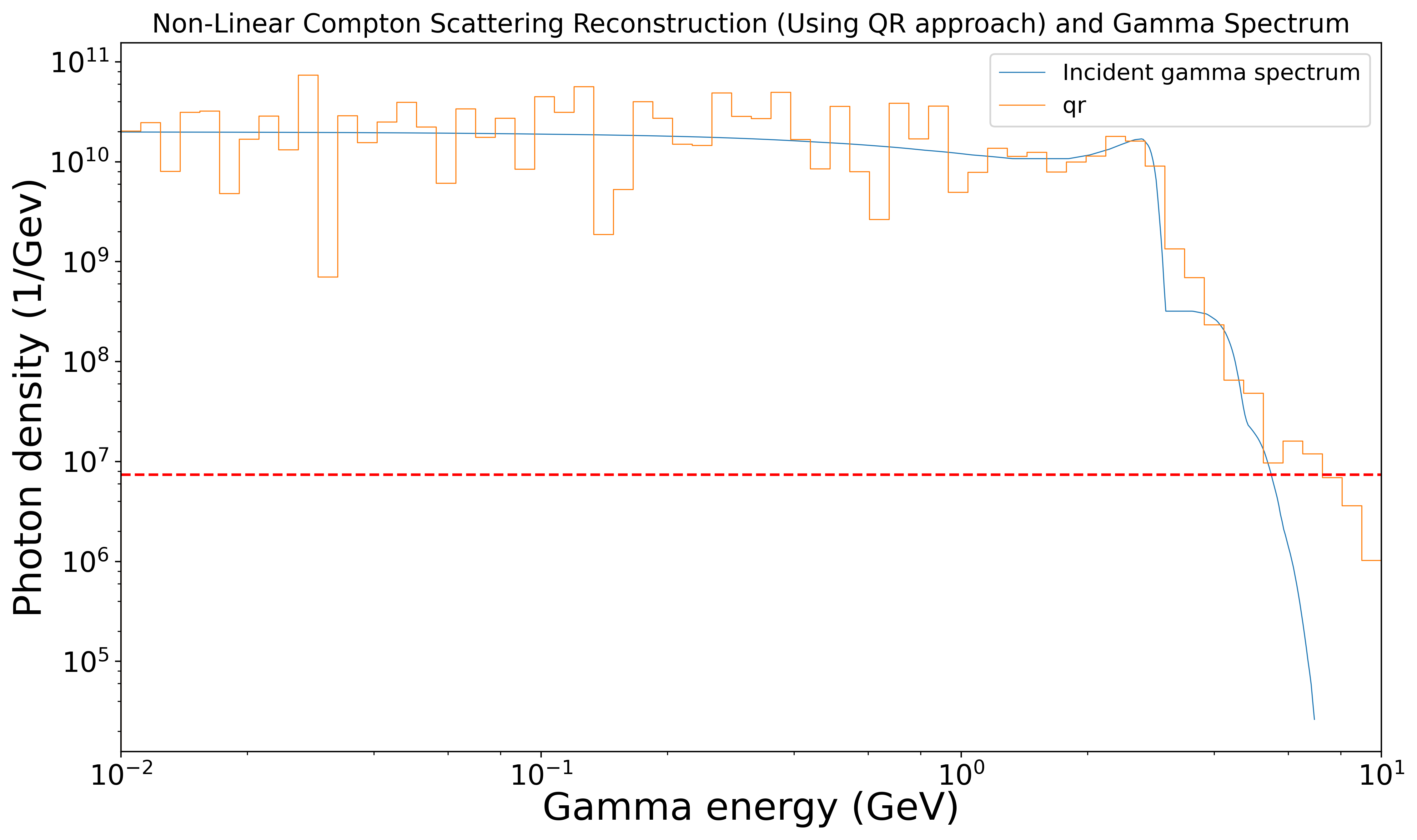}}\\
    \subfloat[Quantum electrodynamics case]{\includegraphics[width=\columnwidth]{qed_QR.png}}\\
    \subfloat[Filamentation case]{\includegraphics[width=\columnwidth]{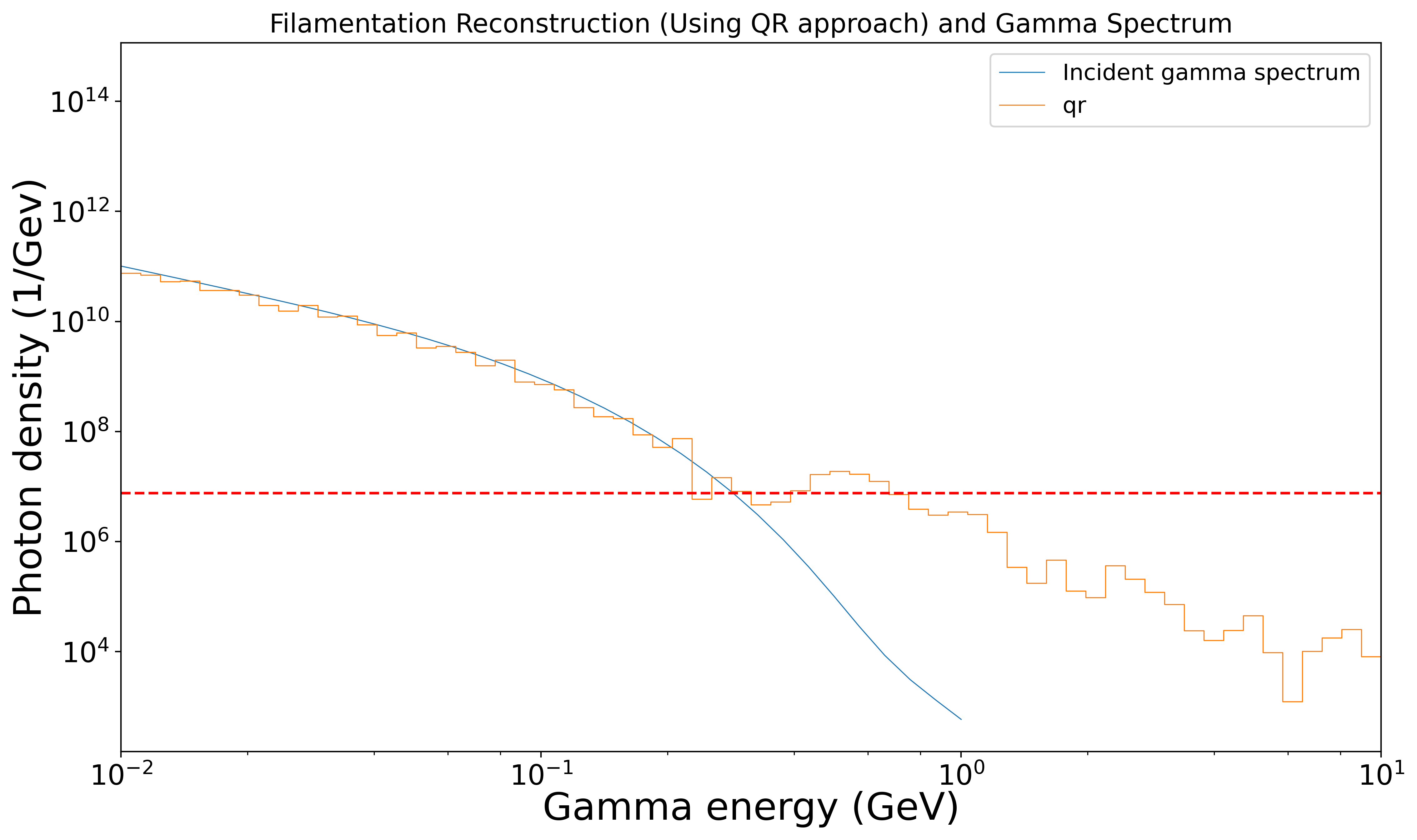}}
  \caption{Smooth reconstructed $\gamma$-ray distributions using QR decomposition.}
  \label{fig:7}
\end{figure}

Examining the predicted distributions for spectra generated from mono-energetic and arbitrary discrete distributions in Fig. \ref{fig:6}, the QR decomposition method predicts the peak of the monoenergetic spectrum well but shows mixed results with the arbitrary case. In the arbitrary discrete case, the reconstruction shows several deviations from the $\gamma$-ray distribution, resulting in predictions that are several orders of magnitude off from the true spectrum. In addition, the noise floor in the reconstruction is extremely low, with frequency predictions below one per bin in the region away from the true distribution.

Across three of the smooth distributions (see Fig. \ref{fig:7}), this method struggles to show any consistent performance. The reconstruction is most accurate in the filamentation case, and it shows mixed results in the other two experimental cases. In NLCS, this reconstruction method shows the most success in the mid GeV range but seems to struggle to find a decent amount of accuracy in the lower energy ranges. A similar pattern of success can be noticed in the quantum electrodynamics case as well.

\subsection{ML Analysis}

To implement an ML analysis, using Eq.~\ref{eq:pedro}, training data was synthesized by creating an arbitrary energy distribution and multiplying each \textit{x}-vector by the matrix to generate the corresponding spectrum (or \textit{y}-vector) that PEDRO would measure. To simulate real-world noise from scattered electrons during the pair production process, low-level noise vectors (each element assigned a number between 0 and $10^4$ selected from a Poisson distribution) were calculated and added to the \textit{y}-vectors.

Once the training and test data sets were synthesized, the model was then provided the set of \textit{y}-vectors and instructed to calculate which \textit{x}-vector interacted with the spectrometer to give rise to such \textit{y}-vector. To determine the optimal settings used for the ML model analysis, including the type of optimizer and number of layers, each model element was varied independently, and the model was trained on the same training data set. Then, the accuracy output from testing the model against the testing data set was used to assess which would produce the most desirable results, with a higher accuracy value indicating the setting was more favorable. 

Table~\ref{table:1} summarizes the model's architecture. The model used the Adam optimizer with a learning rate of 0.005 over 600 epochs. The mean squared error was chosen as the loss function for the model, which was constructed using the Python Keras library \cite{chollet2015keras}.

\begin{table}[!bht]
	\centering
	\caption{A summary of the ML model to predict incoming gamma spectra based on positron-electron detection}	
	\begin{tabular}{ c c c } 
	 \toprule
	 \textbf{Layer} & \textbf{Output} & \textbf{Param} \\ 
	 \textbf{(type, bias, activation)} & \textbf{Shape} & \textbf{Num} \\ 
	 \midrule
	 dense (Dense, true, linear) & (None, 64) & 64 \\ 
	  dense1 (Dense, true, linear) & (None, 64) & 64 \\ 
	\bottomrule
	\end{tabular}
	\label{table:1}
\end{table}
 
Given the established linear relationship between incoming photon energy distributions and the resulting PEDRO spectra, a decision was taken to incorporate only linear activation functions in the model. The number of layers, the output shape, and the number of parameters were all tested through a series of trials. The above configuration construction was determined to have the best predictive power. 

\subsubsection{Assessing the method}

\begin{figure}[hbt!]
  \centering
    \subfloat[Mono-energetic case]{\includegraphics[width=0.47\textwidth]{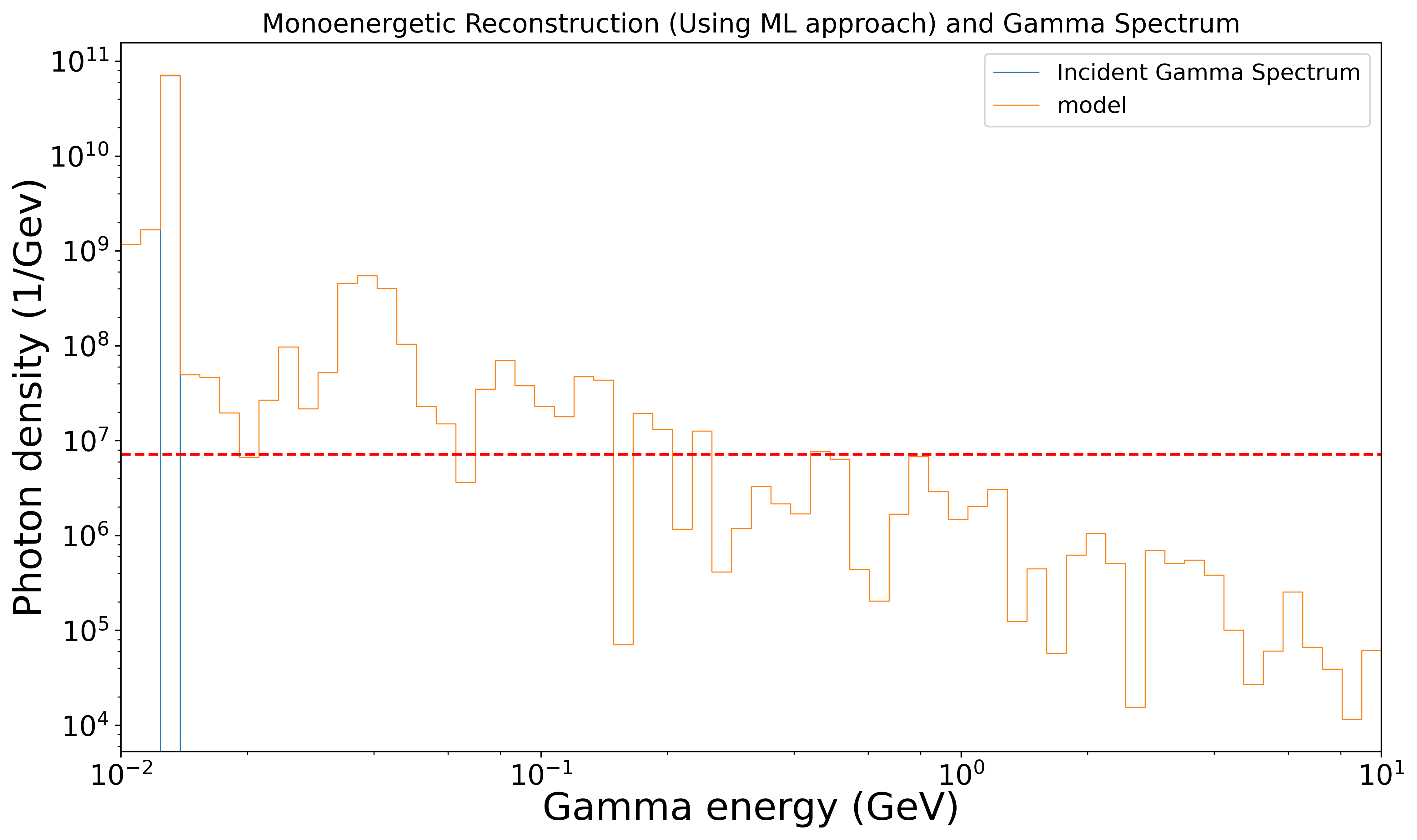}}\\
    \subfloat[Arbitrary Discrete case]{\includegraphics[width=0.47\textwidth]{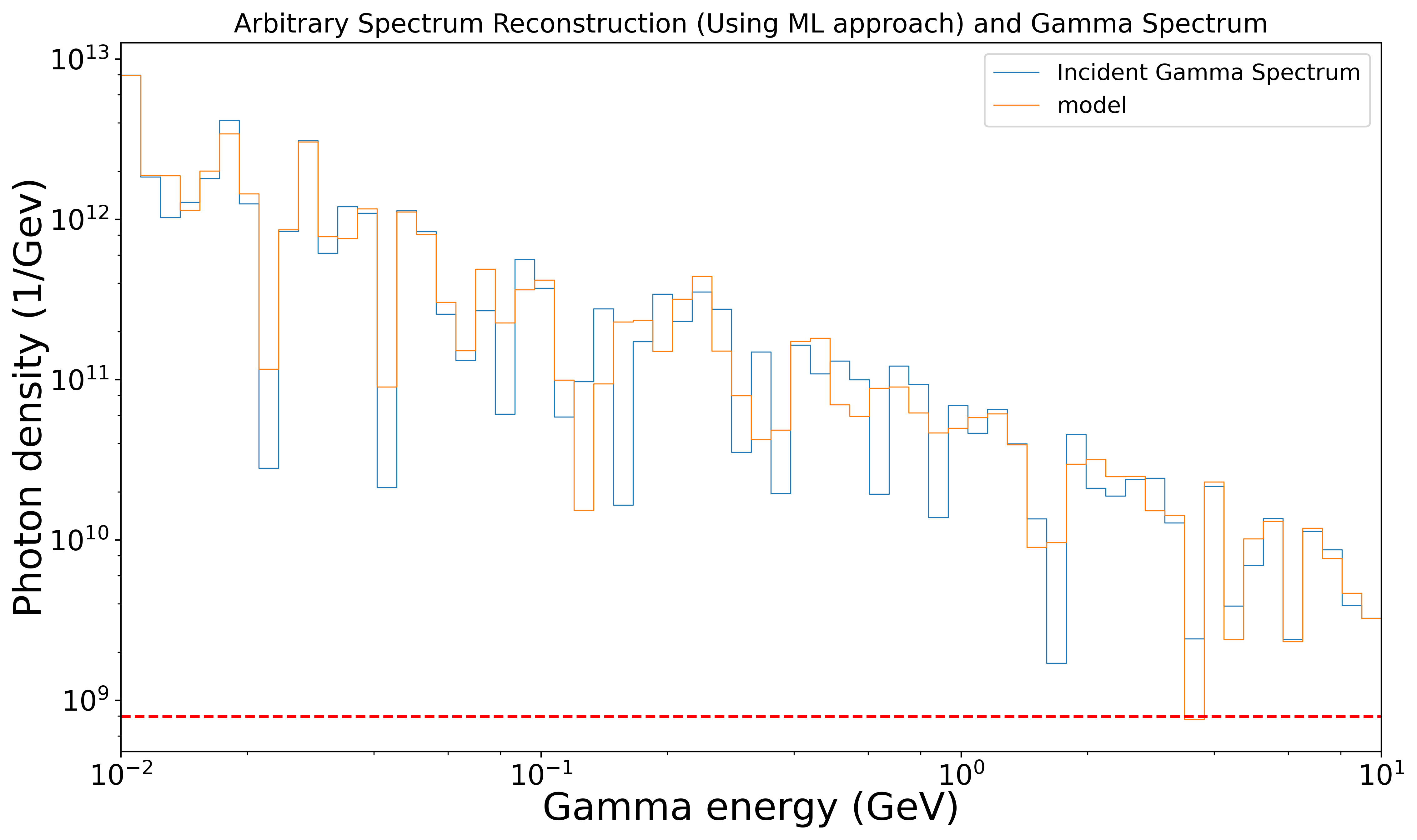}}\\
  \caption{Discrete reconstructed gamma energy distributions using ML.}
  \label{fig:2}
  \vspace*{-\baselineskip}
\end{figure}

Using ML, reconstruction success varies depending on the individual cases. Examining the discrete cases in Fig. \ref{fig:2}, the trained ML model adeptly predicts peaks in the correct bins corresponding to the photon energies in the mono-energetic case. The scale is also accurately reconstructed, followed by predictions of photon frequencies several orders of magnitude lower than the incident peak.
 
In the arbitrary discrete case, the model effectively reconstructs the distribution with the most deviations visually apparent in between bins 30 and 40 (approximately corresponding to an energy range 250-670 MeV), as seen in Fig. \ref{fig:2}(b). Despite these variations, the model reconstruction falls within one order of magnitude across the entire spectrum, indicating its effectiveness at all energy ranges.

\begin{figure}[hbt!]
  \centering
  \subfloat[Nonlinear Compton scattering case]{\includegraphics[width=\columnwidth]{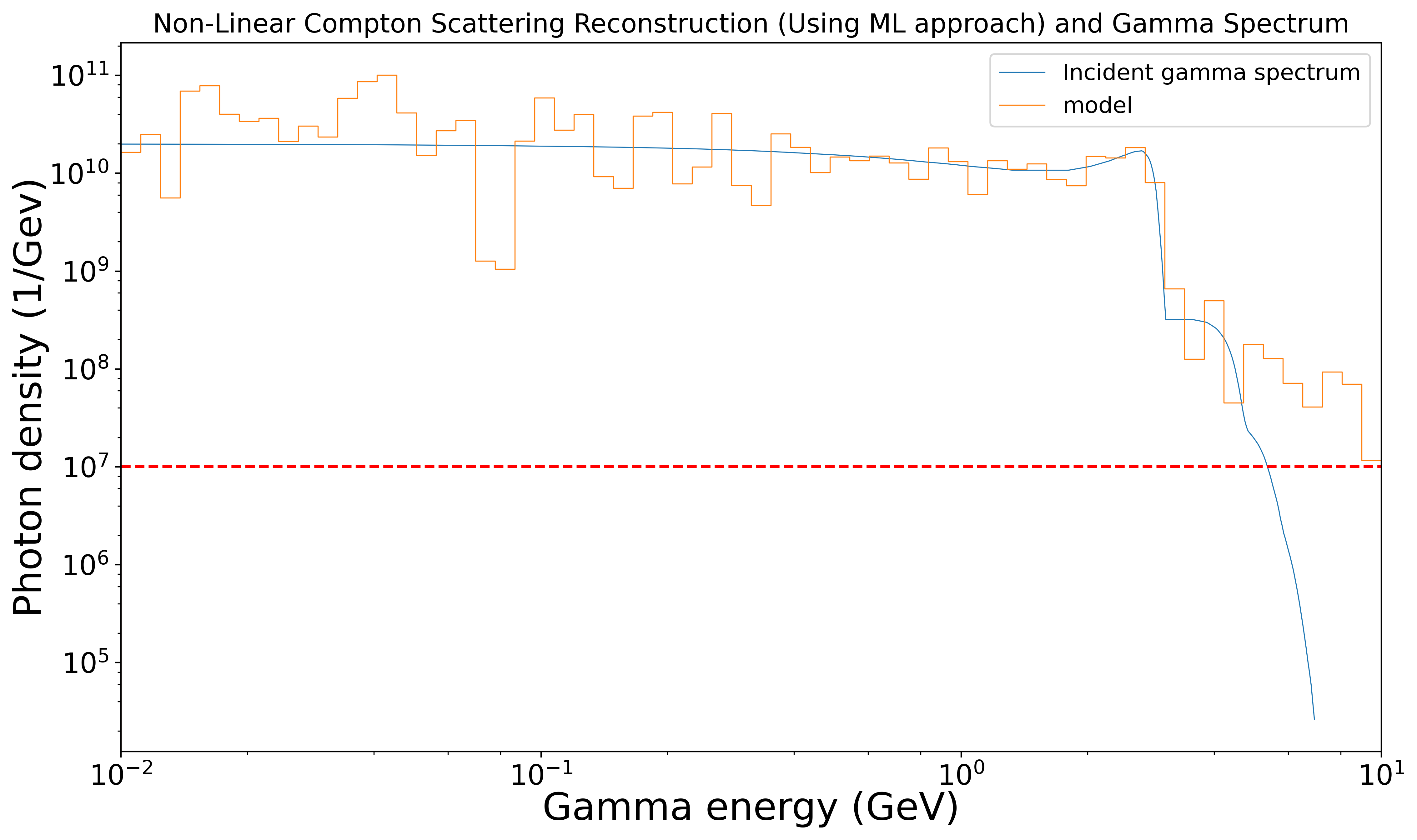}}\\ 
  \subfloat[Quantum electrodynamics case]{\includegraphics[width=\columnwidth]{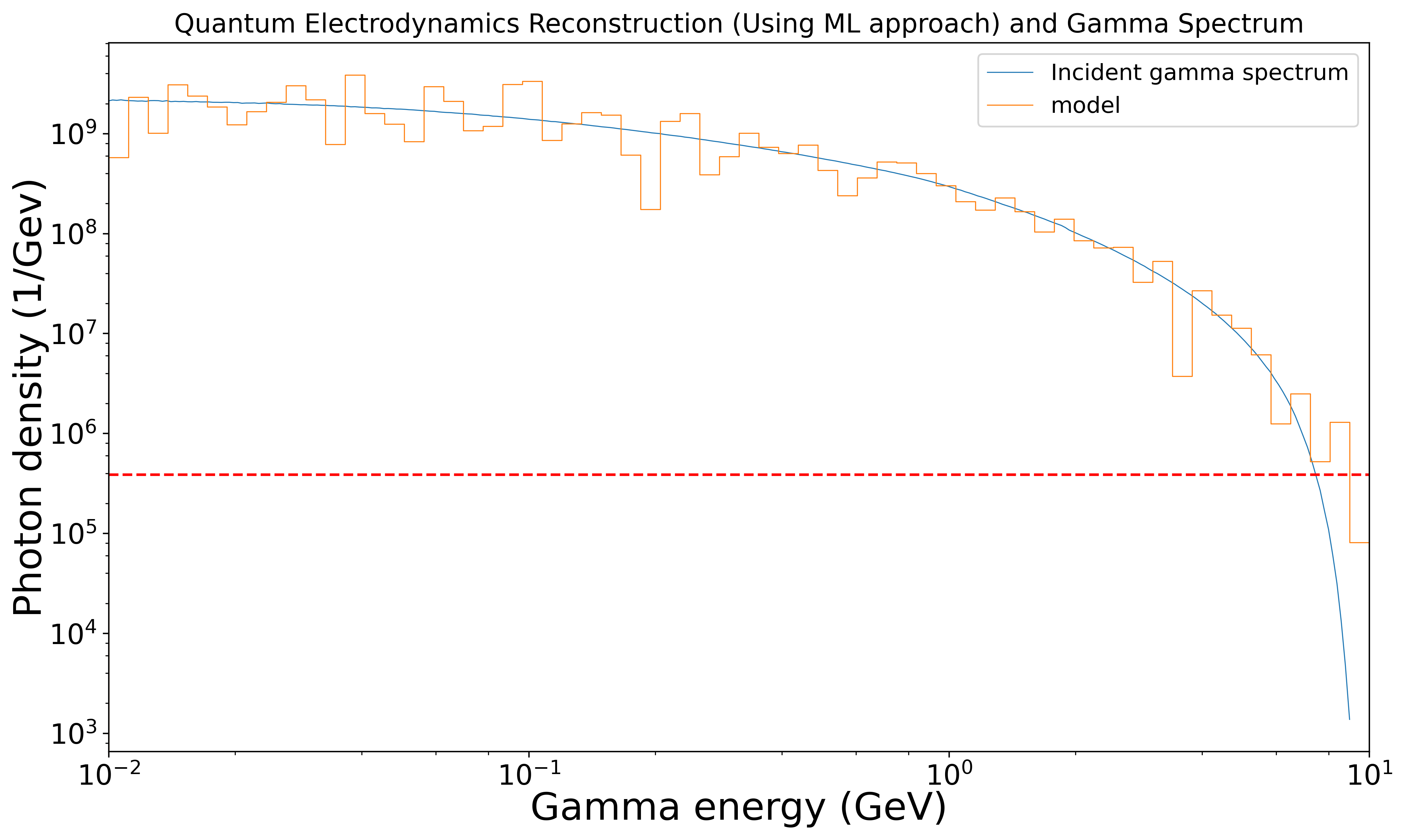}}\\
  \subfloat[Filamentation case]{\includegraphics[width=\columnwidth]{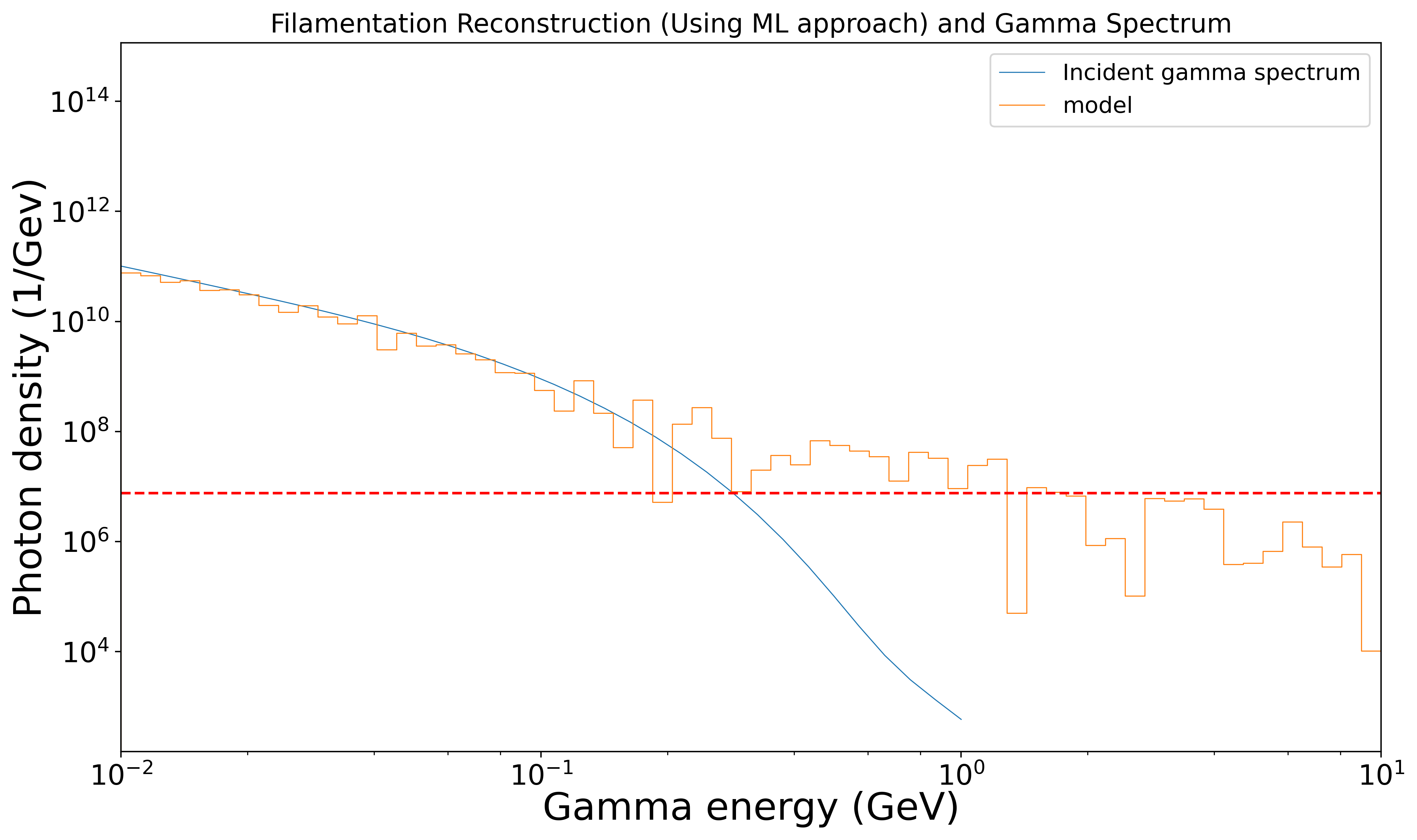}}
  \caption{Smooth reconstructed gamma distributions using ML.}
  \label{fig:3}
\end{figure}

Figure \ref{fig:3} summarizes the model's performance in smooth cases that more closely represent experimental results. In Fig. \ref{fig:3}, the model predicts the correct distribution pattern between bins 1 and 30 (approximately corresponding to a range of 10-250 MeV). In the higher energy bins, the model's predicted photons fall within the spectrometer's dynamic range in the nonlinear Compton scattering (Fig. \ref{fig:3}(a)) and quantum electrodynamics (Fig. \ref{fig:3}(b)) cases. In these two cases, the model follows the entire pattern of the reconstructed distribution, consistently staying within the same order of magnitude as the $\gamma$-ray distribution except at the highest energies, which are not of high physical interest.

In the filamentation case Fig. \ref{fig:3}(c), the model is aligned with the $\gamma$-ray distribution in the 10-250 MeV range. Though the deviations grow larger as the energy increases, as far as the dynamic range is concerned, the machine learning model performs within an acceptable level of accuracy. 

\subsection{Maximum Likelihood Estimation and ML (Hybrid)}

As noted above, MLE is an algorithm that, given an initial guess for the solution of an equation, iteratively converges to the nearest solution that is the most probable. In the current context, this approach transforms the task of deducing the original energy distribution from the PEDRO output. Rather than approaching it as an analytical problem, MLE treats it as a statistical challenge aimed at estimating a set of parameters \cite{MYUNG200390}. 

While the MLE algorithm does not require training like the ML model, it does require an initial guess that is sufficiently close to the proper solution to Eq. \ref{eq:pedro}. Given the enormous potential for variety in \textit{y}-vectors and incoming energy distributions, a random guess may not reliably converge to the desired result. However, given the utility of ML, it is possible to provide a customized guess for every \textit{y}-vector: a guess provided by the ML model. This combination serves as a hybrid approach to recovering the original energy distribution instead of pure ML (or QR decomposition, as discussed above). Figure \ref{fig:flowchart} demonstrates the implementation of this method.

\tikzstyle{block} = [rectangle, draw, fill=blue!20, 
    text width=5em, text centered, rounded corners, minimum height=4em]
\tikzstyle{cloud} = [draw, ellipse,fill=red!20, text width=5em, text centered, node distance=3cm, minimum height=4em]
\tikzstyle{line} = [draw, -latex']

\begin{figure}[hbt!]   
    \begin{center}
        \begin{tikzpicture}[node distance = 3 cm, auto]
            \node [block] (x-vec) {Incoming $\gamma$};
            \node [block, right of = x-vec] (R) {Pair production from wire};
            \node [block, right of = R] (y-vec) {PEDRO Spectrum};
            \node [block, below of = y-vec ] (init) {ML model guess};
            \node [block, left of=init] (soln) {MLE smoothing};
            \node [block, left of=soln] (end) {Final Solution};
            \path [line] (x-vec) -- (R);
            \path [line] (R) -- (y-vec);
            \path [line] (y-vec) -- (init);
            \path [line] (init) -- (soln);
            \path [line] (soln) -- (end);
        \end{tikzpicture}
    \end{center}   
    \caption{This flowchart illustrates how data from PEDRO will be processed in the hybrid approach.}
    \label{fig:flowchart}
\end{figure}
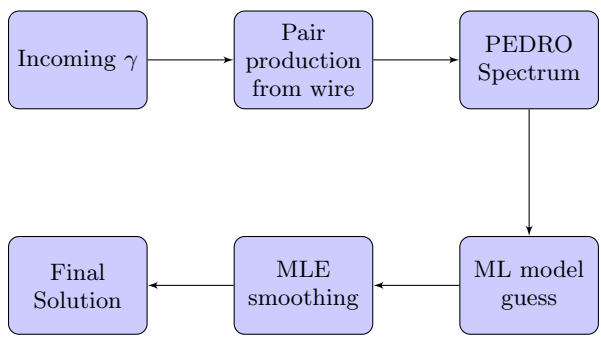

Using the ML model's guess, which may be close to the true distribution, the MLE algorithm may also reliably converge to the true distribution. This is particularly advantageous as it mitigates the challenge of converging to incorrect points.

\subsubsection{Assessing the method}

\begin{figure}[hbt!]
  \centering
   \subfloat[Mono-energetic case]{\includegraphics[width=\columnwidth]{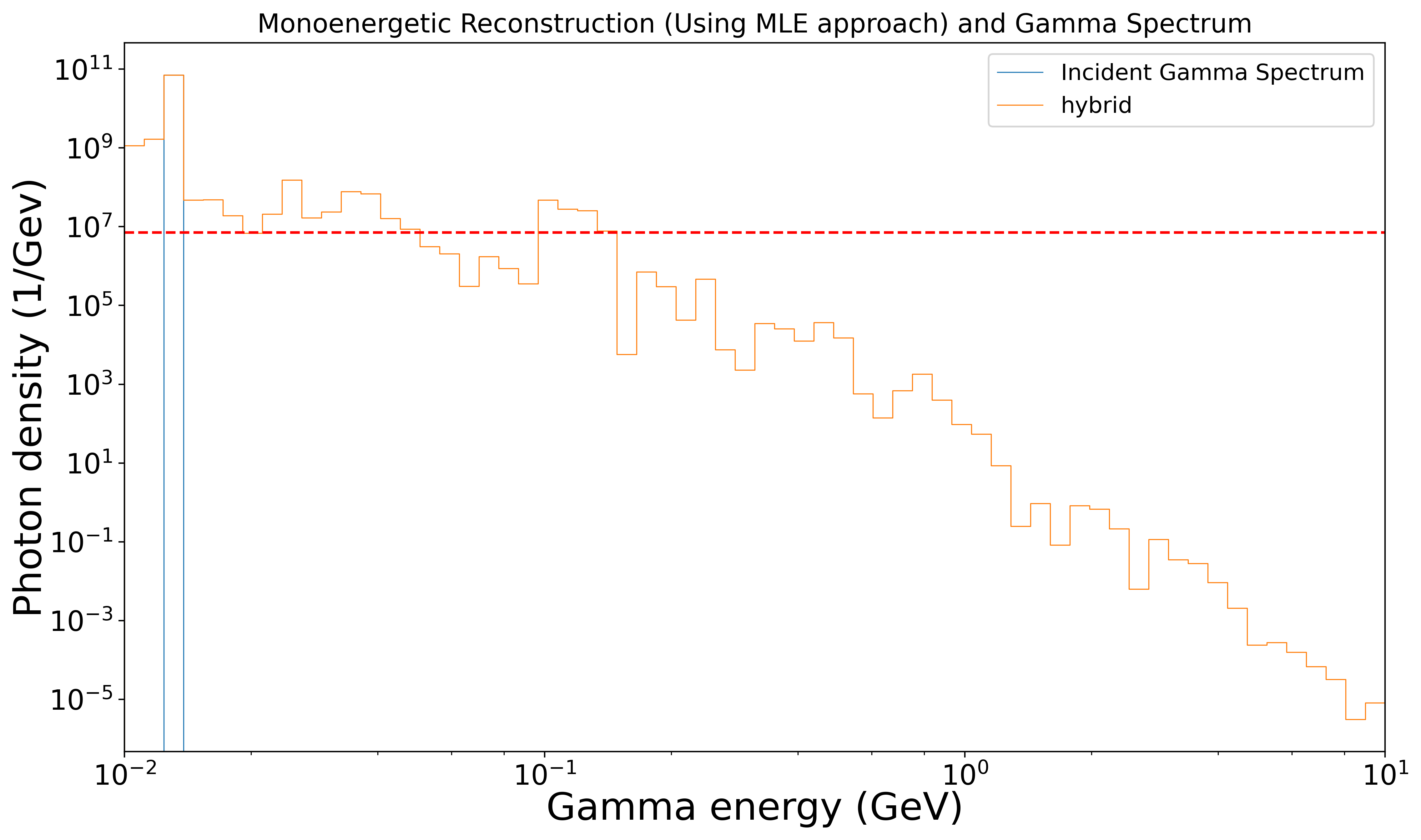}}\\
    \subfloat[Arbitrary discrete case]{\includegraphics[width=\columnwidth]{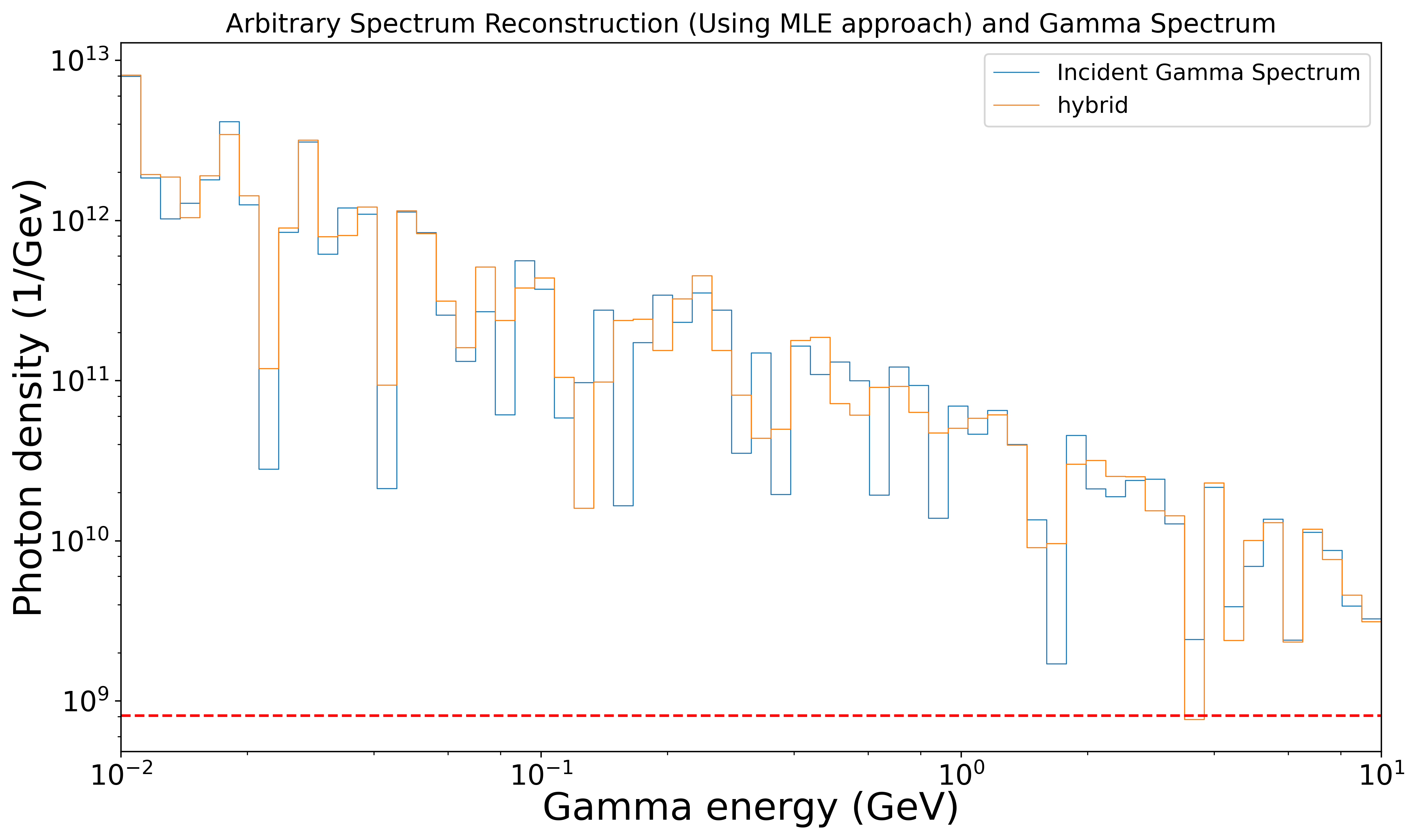}}
  \caption{Discrete reconstructed gamma energy distributions using the hybrid approach.}
  \label{fig:4}
\end{figure}

In the discrete monoenergetic case Fig. \ref{fig:4}, the hybrid approach correctly predicts the location and heights of the peak associated with the energy ranges of the true photon spectra. Since the red line indicates the dynamic range, the reconstruction shows that most of the significant photon counts are located at the exact mono-energetic peak, while the rest of the bins have negligible amounts of photons.

In the arbitrary case, there are certain ranges where the hybrid approach does very accurately re-create the original spectrum, while in other areas it has less accuracy but within an order of magnitude. In particular, the most deviation seems to occur within the 100 MeV to 1 GeV range.

\begin{figure}[hbt!]
  \centering
    \subfloat[Nonlinear Compton scattering case]{\includegraphics[width=\columnwidth]{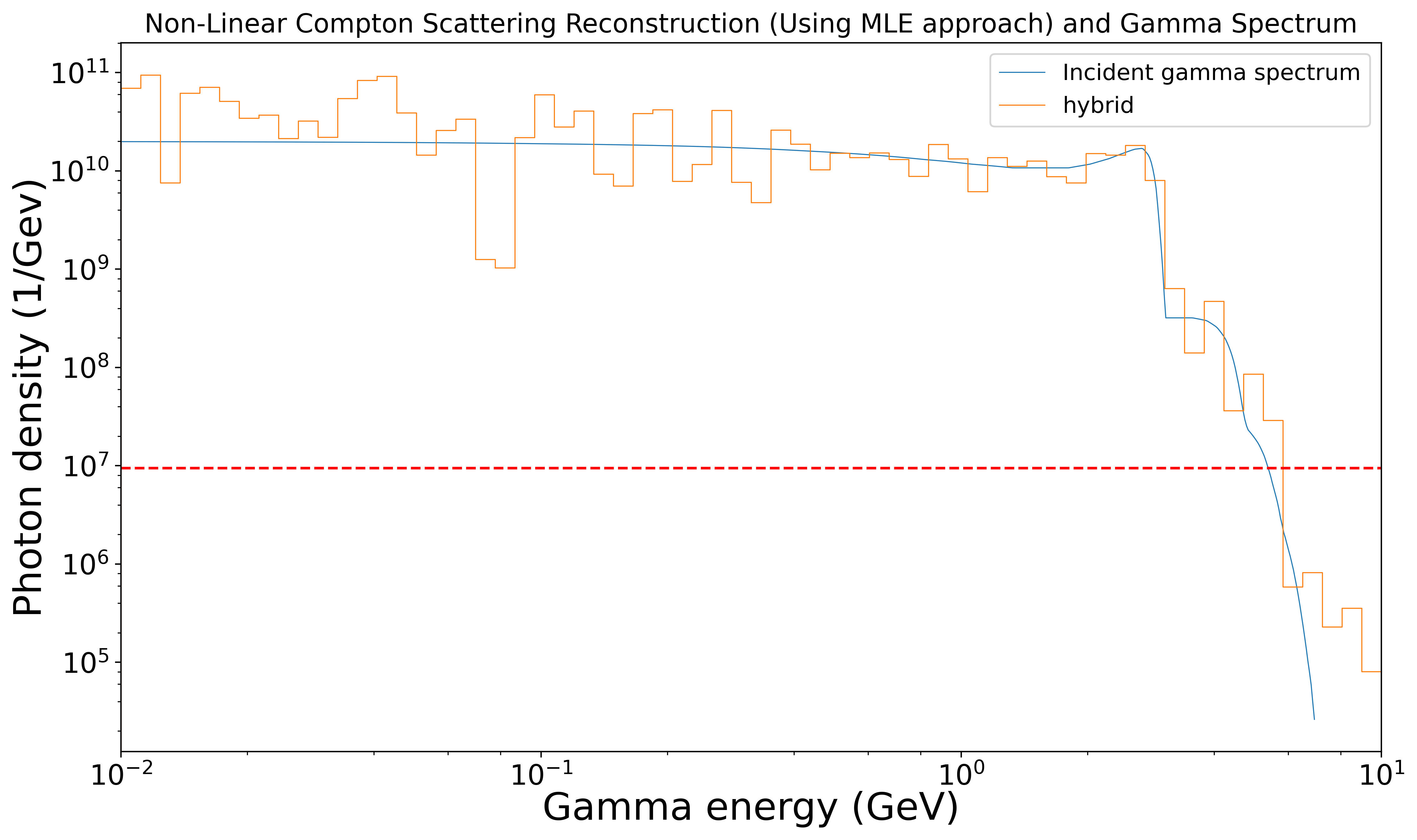}}\\
    \subfloat[Quantum electrodynamics case]{\includegraphics[width=\columnwidth]{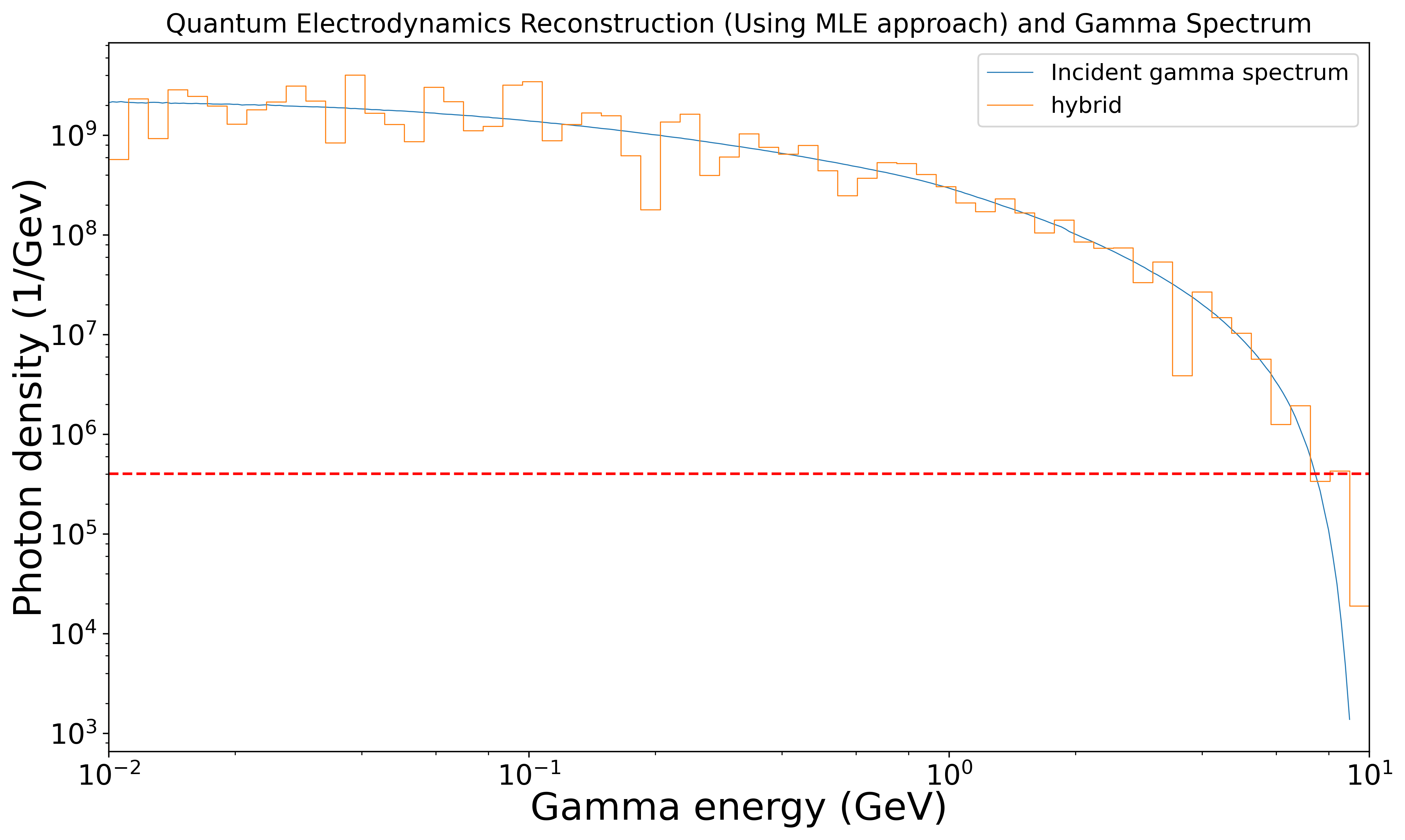}}\\
    \subfloat[Filamentation case]{\includegraphics[width=\columnwidth]{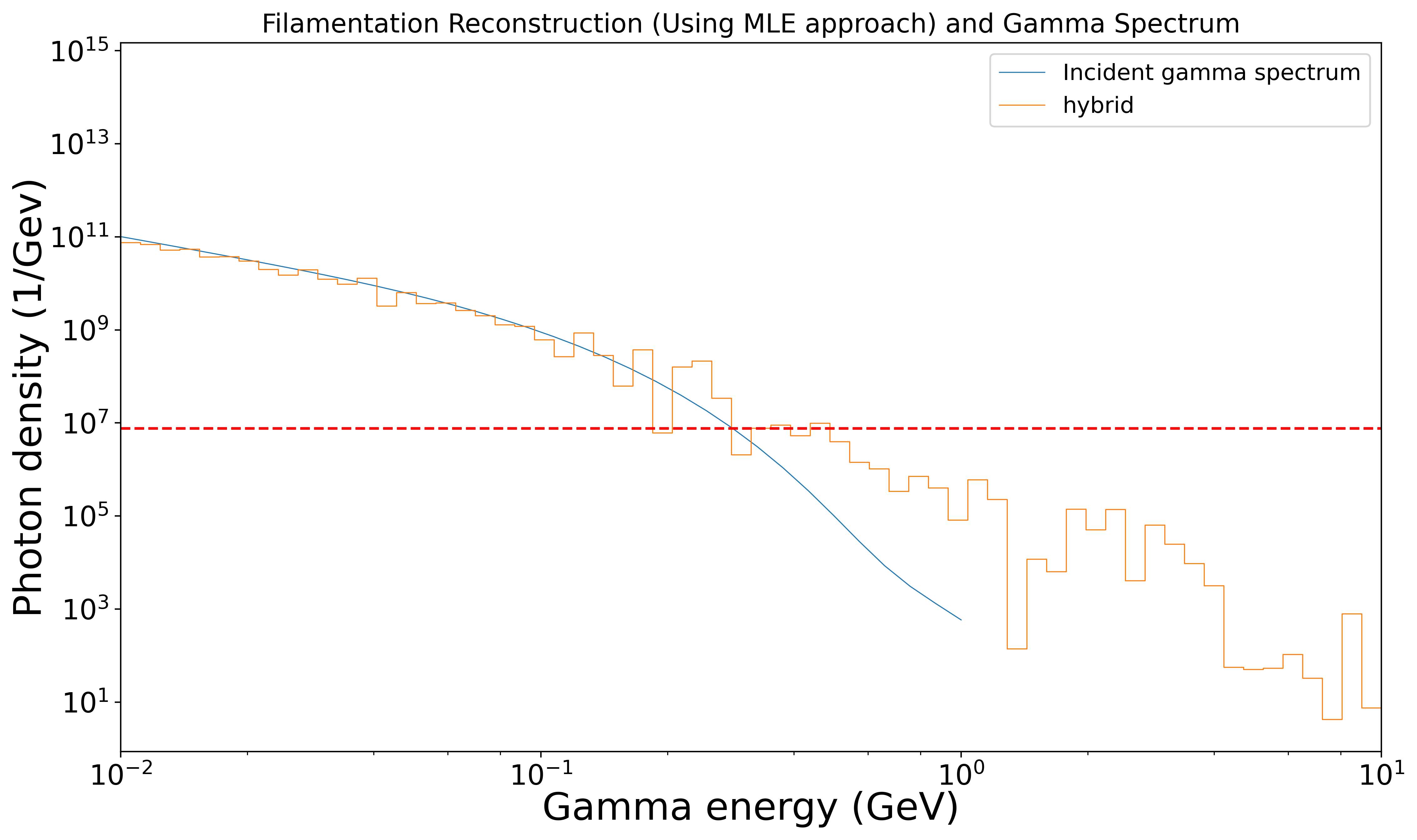}}
  \caption{Smooth reconstructed $\gamma$-ray distributions using the hybrid approach.}
  \label{fig:5}
\end{figure}
 
In the smooth cases from Fig.~\ref{fig:5}, the hybrid approach closely follows the pattern of the $\gamma$-ray distribution across all of the energy ranges (with a deviation from the filamentation case above 250 MeV). In the nonlinear Compton scattering and quantum electrodynamics cases, the reconstructions consistently fall within the same order of magnitude as the $\gamma$-ray distribution. 

Compared to the ML model's performance, the hybrid approach regularly fixes some deviations and smoothens the reconstruction, improving the accuracy. Additionally, it seems that the reconstruction is more effective in the higher energy ranges for quantum electro dynamics and NLCS, while it seems to be the most effective the lower energy ranges for filamentation. Regardless, it has a relatively consistent reconstruction accuracy across the spectrum for all three cases when looking at within the dynamic range.

\section{Discussion}
\label{sec:discussion}

\begin{figure}[hbt!]
    \centering \includegraphics[width=\columnwidth]{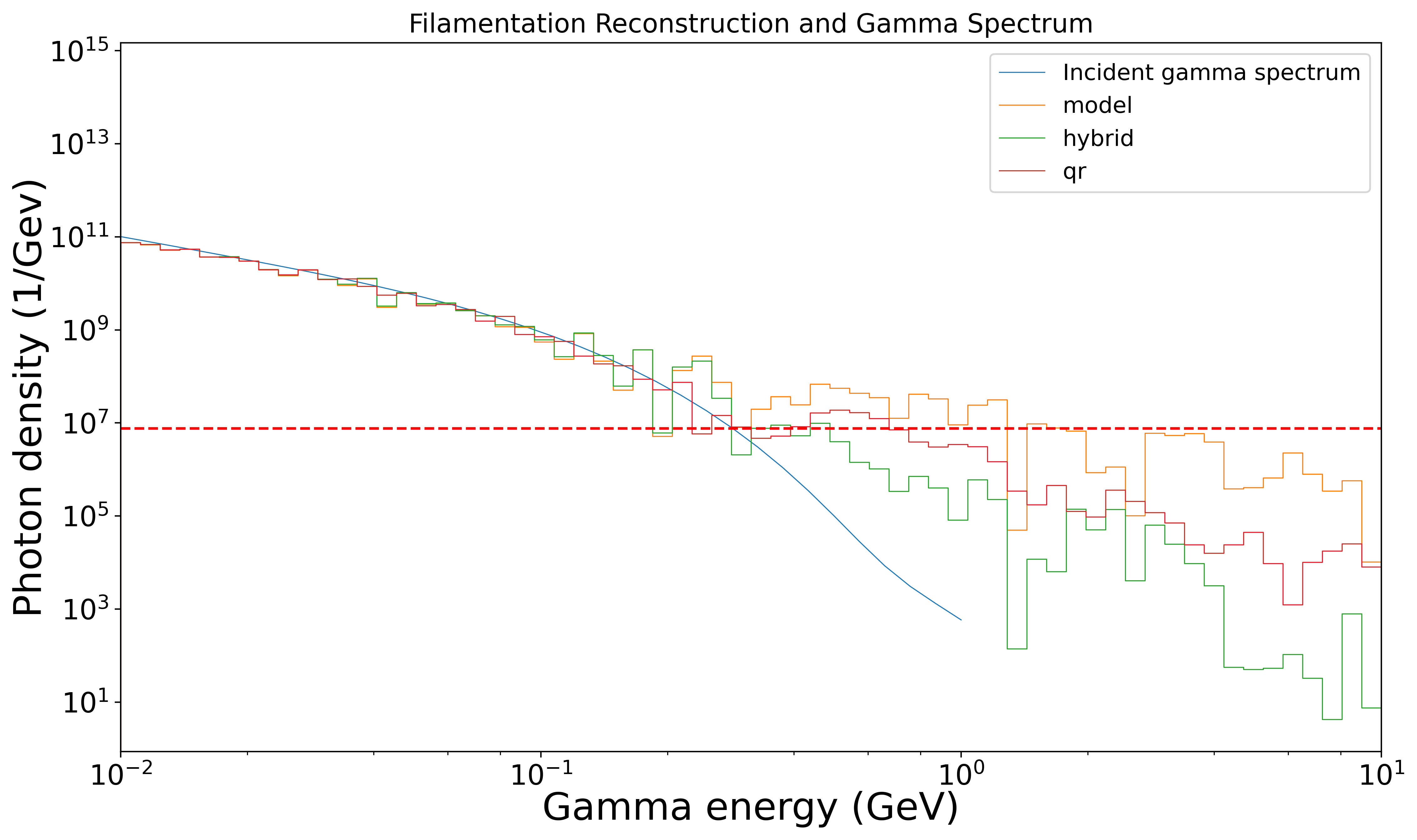}   \caption{Reconstruction of filamentation case using three different methods.}
    \label{fig:All}
\end{figure}

In each of the experimental cases, the ML and hybrid approaches seem to yield a similar level of accuracy and precision during the reconstruction. For a side-by-side comparison, the performances of each approach in the filamentation case were plotted on a single graph and presented in Fig. \ref{fig:All}. When looking at the performance of each of these approaches, they are practically indistinguishable within the dynamic range, as indicated by the red dotted line as the lower limit.

However, when examining the quantum electrodynamics case, it becomes much more evident that the QR decomposition method performs worse than the hybrid or ML approaches. This suggests that when looking across multiple real world simulations, the hybrid approach or the ML approach should be considered as the  candidate for reconstruction.

 \begin{figure}[hbt!]
    \centering \includegraphics[width=\columnwidth]{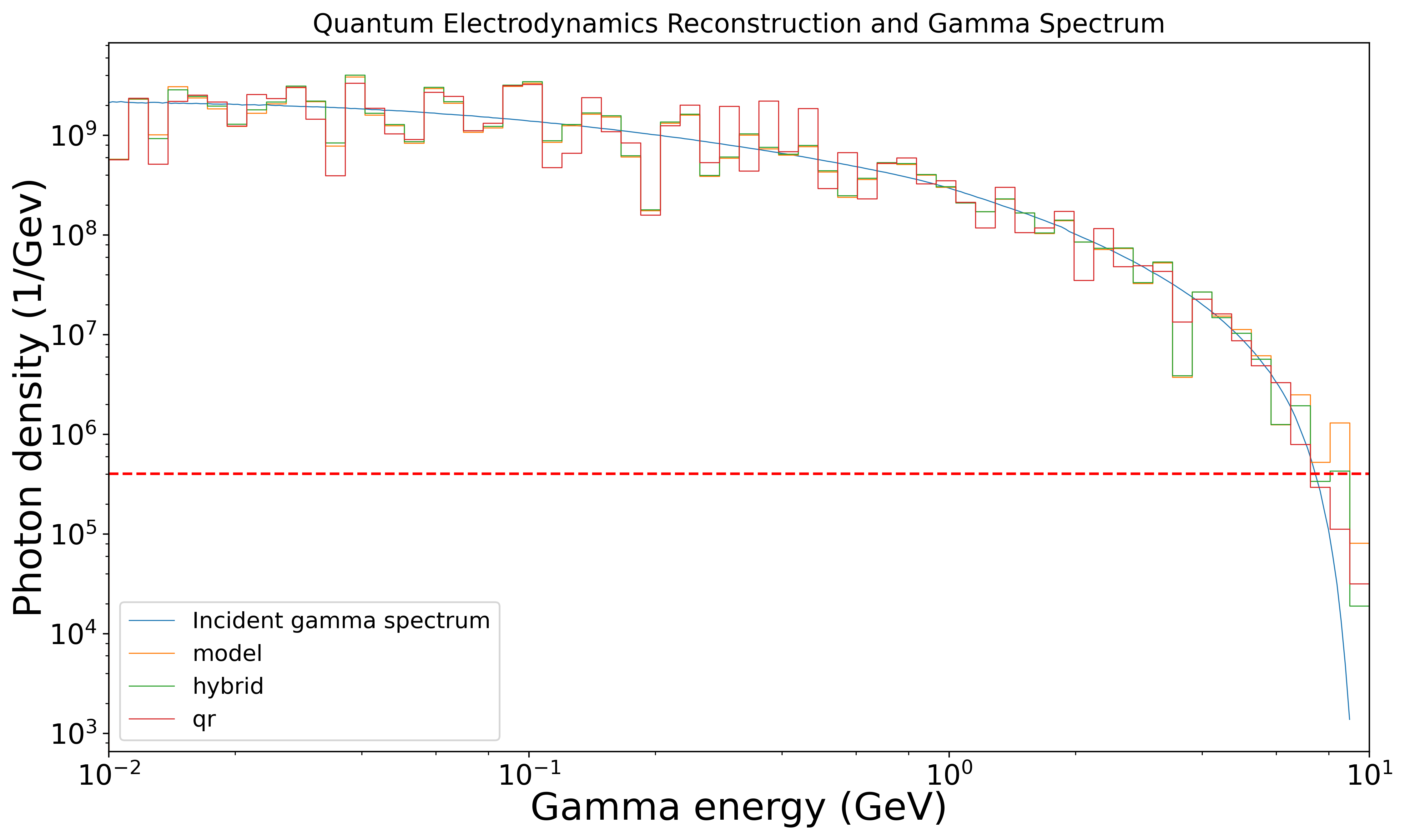}   \caption{Reconstruction of quantum electrodynamics case using three different methods.}
    \label{fig:All_QED}
\end{figure}

In terms of computational time and maintenance, the QR decomposition method proved to be the fastest, but it was also very much the least reliable. As noted in Sec. \ref{QR}, QR decomposition generates varying levels of accuracy based on the experimental case that was analyzed. This means that this method should not be used in real world scenarios because its precision will vary based on the experimental case. The experimental noise, as demonstrated in Sec. \ref{QR_compare}, also throws the method off balance, thus making it the least reliable. 

While steps are continually being taken on the instrument side to reduce noise issues, any reconstruction method should remain robust in the face of noise and should remain consistent in its performance regardless of the experimental case presented. Given these requirements, it is still likely that the hybrid approach will prove to be the most effective in reconstructing $\gamma$-ray spectra from PEDRO readings\cite{yadav2021_AAC, 2021yadav,Naranjo_1}.

\section{Outlook and future work}
\label{sec:conclusion}

This paper focused on analyzing a noisy linear system with multiple candidates for reconstruction, utilizing photon-strong nuclear field interactions to model PEDRO's measurements of arbitrary and experimental $\gamma$-ray distributions. Comparing QR decomposition, a neural network, and a neural network-informed guess for MLE demonstrated clearly that QR decomposition failed to hold up to noise, a problem the other approaches tackled more effectively. In all experimental cases, the hybrid approach of using ML to generate an informed initial guess for iterative MLE proved to be the most efficient and robust to noise. These results support the implementation of this hybrid approach in other aspects of accelerator physics analysis. 

This work provides support for the larger implementation of combined ML-MLE approaches to the reconstruction of beam parameters in high-energy physics. Using ML to even accomplish a rough reconstruction of any target parameter can be paired with MLE, an algorithm that can serve to hone in on the true solution with much higher confidence of avoiding any pathological solutions. Because the algorithm itself is immune to noise, neural networks can be used to eliminate distortions and other confounding variables as much as possible before refining the solution. 

The work done with PEDRO assumed the data to be analyzed derived from the single-shot case, where only one measurement would be taken for a given electron beam-laser interaction. However, in the multi-shot case, the same interaction can be repeatedly measured with different horizontal positions of the converter target wired to obtain double-differential spectra, yielding more information about these high-energy interactions. Moreover, we may look at sparsely populated regions of the spectrum, thus extending the dynamic range obtained. As such, a future direction for the work in the paper is to generalize its reconstruction methods to the more complicated but rewarding multi-shot case.

These physics goals discussed above are all found in the wide-ranging FACET-II program and other present and future laboratory facilities. As radiation-based diagnostics are key, robust signatures of the physics involved in these experiments, the introduction of modern methods for analyzing such data is an essential component of the FACET-II effort. This paper has provided a data analysis study for the advanced gamma spectroscopy instrumentation at FACET-II. It shows that these methods, in tandem with advanced spectrometers, are promising for extracting the needed physics insights from experiments ranging from PWFA to strong field QED.

\section{Acknowledgements}

This work was performed with the support of the US Department of Energy, Division of High Energy Physics, under Contract No. DE-SC0009914, NSF PHY-1549132 Center from Bright Beams, DARPA under Contract N.HR001120C007  and the STFC Liverpool Centre for Doctoral Training on Data Intensive Science (LIV.DAT) under grant agreement ST/P006752/1. This work used computational and storage services associated with the SCARF cluster, provided by the STFC Scientific Computing Department, United Kingdom.

\bibliography{main.bib}

\end{document}